\documentclass[twocolumn]{aastex63}
\usepackage{float}
\usepackage{graphicx}
\usepackage{amsmath}
\usepackage{natbib}
\usepackage{color}
\usepackage{verbatim}
\usepackage[utf8]{inputenc}
\usepackage{mathtools}
\usepackage{amssymb}
\usepackage{textcomp}
\usepackage{enumitem}
\usepackage{braket}

\newcommand{\magsqarc}{mag\,arcsec\,$^{-2}$}

\renewcommand{\thetable}{\arabic{table}}

\received{31 October, 2019}
\revised{31 August, 2020}
\accepted{23 October, 2020}
\submitjournal{\textsc{The Astrophysical Journal}}
\shortauthors{Smercina \textit{et al.}}

\begin{document}
\title{The Saga of M81: Global View of a Massive Stellar Halo in Formation}

\author[0000-0003-2599-7524]{Adam Smercina}
\affiliation{Department of Astronomy, University of Michigan, Ann Arbor, MI 48109, USA}
\affiliation{Astronomy Department, University of Washington, Box 351580, Seattle, WA 98195-1580, USA}
\author[0000-0002-5564-9873]{Eric F. Bell}
\affiliation{Department of Astronomy, University of Michigan, Ann Arbor, MI 48109, USA}
\author{Paul A. Price}
\affiliation{Department of Astrophysical Sciences, Princeton University, Princeton, NJ 08544, USA}
\author[0000-0002-0558-0521]{Colin T. Slater}
\affiliation{Astronomy Department, University of Washington, Box 351580, Seattle, WA 98195-1580, USA}
\author{Richard D'Souza}
\affiliation{Department of Astronomy, University of Michigan, Ann Arbor, MI 48109, USA}
\affiliation{Vatican Observatory, Specola Vaticana, V-00120, Vatican City State} 
\author[0000-0001-6380-010X]{Jeremy Bailin}
\affiliation{Department of Physics and Astronomy, University of Alabama, Box 870324, Tuscaloosa, AL 35487-0324, USA}
\author[0000-0001-6982-4081]{Roelof S. de Jong}
\affiliation{Leibniz-Institut f\"{u}r Astrophysik Potsdam (AIP), An der Sternwarte 16, 14482 Potsdam, Germany}
\author[0000-0002-2502-0070]{In Sung Jang}
\affiliation{Leibniz-Institut f\"{u}r Astrophysik Potsdam (AIP), An der Sternwarte 16, 14482 Potsdam, Germany}
\author[0000-0003-2325-9616]{Antonela Monachesi}
\affiliation{Instituto de Investigaci\'on Multidisciplinar en Ciencia y Tecnolog\'ia, Universidad de La Serena, Ra\'ul Bitr\'an 1305, La Serena, Chile}
\affiliation{Departamento de F\'isica y Astronom\'ia, Universidad de La Serena, Av. Juan Cisternas 1200 N, La Serena, Chile}
\author[0000-0002-1793-3689]{David Nidever}
\affiliation{Department of Physics, Montana State University, P.O. Box 173840, Bozeman, MT 59717-3840}
\affiliation{National Optical Astronomy Observatory, 950 North Cherry Ave, Tucson, AZ 85719}

\email{asmerci@umich.edu}

\begin{abstract}
Recent work has shown that Milky Way-mass galaxies display an incredible range of stellar halo properties, yet the origin of this diversity is unclear. The nearby galaxy M81 --- currently interacting with M82 and NGC 3077 --- sheds unique light on this problem. We present a Subaru Hyper Suprime-Cam survey of the resolved stellar populations around M81, revealing M81's stellar halo in never-before-seen detail. We resolve the halo to unprecedented $V$-band equivalent surface brightnesses of 33\,\magsqarc, and produce the first-ever global stellar mass density map for a Milky Way-mass stellar halo outside of the Local Group. Using the minor axis, we confirm M81's halo as one of the lowest mass and metal-poorest known ($M_{\star} \simeq 1.16{\times}10^9 M_{\odot}$, [Fe/H] $\simeq {-}1.2$) --- indicating a relatively quiet prior accretion history. Yet, our global halo census finds that tidally unbound material from M82 and NGC 3077 provides a substantial infusion of metal-rich material ($M_{\star} \simeq 5.4{\times}10^8$ $M_{\odot}$, [Fe/H] $\simeq {-}$0.9). We further show that, following the accretion of its massive satellite M82 (and the LMC-like NGC 3077), M81 will host one of the most massive and metal-rich stellar halos in the nearby universe. Thus, the saga of M81: following a passive history, M81's merger with M82 will completely transform its halo from a low-mass, anemic halo rivaling the MW, to a metal-rich behemoth rivaled only by systems such as M31. This dramatic transformation indicates that the observed diversity in stellar halo properties is primarily driven by diversity in the largest mergers these galaxies have experienced. \\
\end{abstract}

\section{Introduction}
\label{sec:intro}
In the $\Lambda$--Cold Dark Matter ($\Lambda$CDM) paradigm, galaxies assemble hierarchically, experiencing frequent mergers with other galaxies \citep[e.g.,][]{white&rees1978,bullock2001}. These events transform the morphological and kinematic structure of the central galaxy \citep{toomre&toomre1972}, and funnel cold gas into the center of the gravitational potential, stimulating the formation of new generations of stars and enriching the existing interstellar reservoirs \citep{barnes&hernquist1991}. As a result of short ($\lesssim$1\,Gyr) dynamical and star formation timescales, the impacts of such mergers quickly become well-mixed into the main body of the galaxy, making it incredibly difficult to infer the properties of the progenitor merging system long afterwards.

\begin{figure*}[t]
\leavevmode
\centering
\includegraphics[width={0.85\linewidth}]{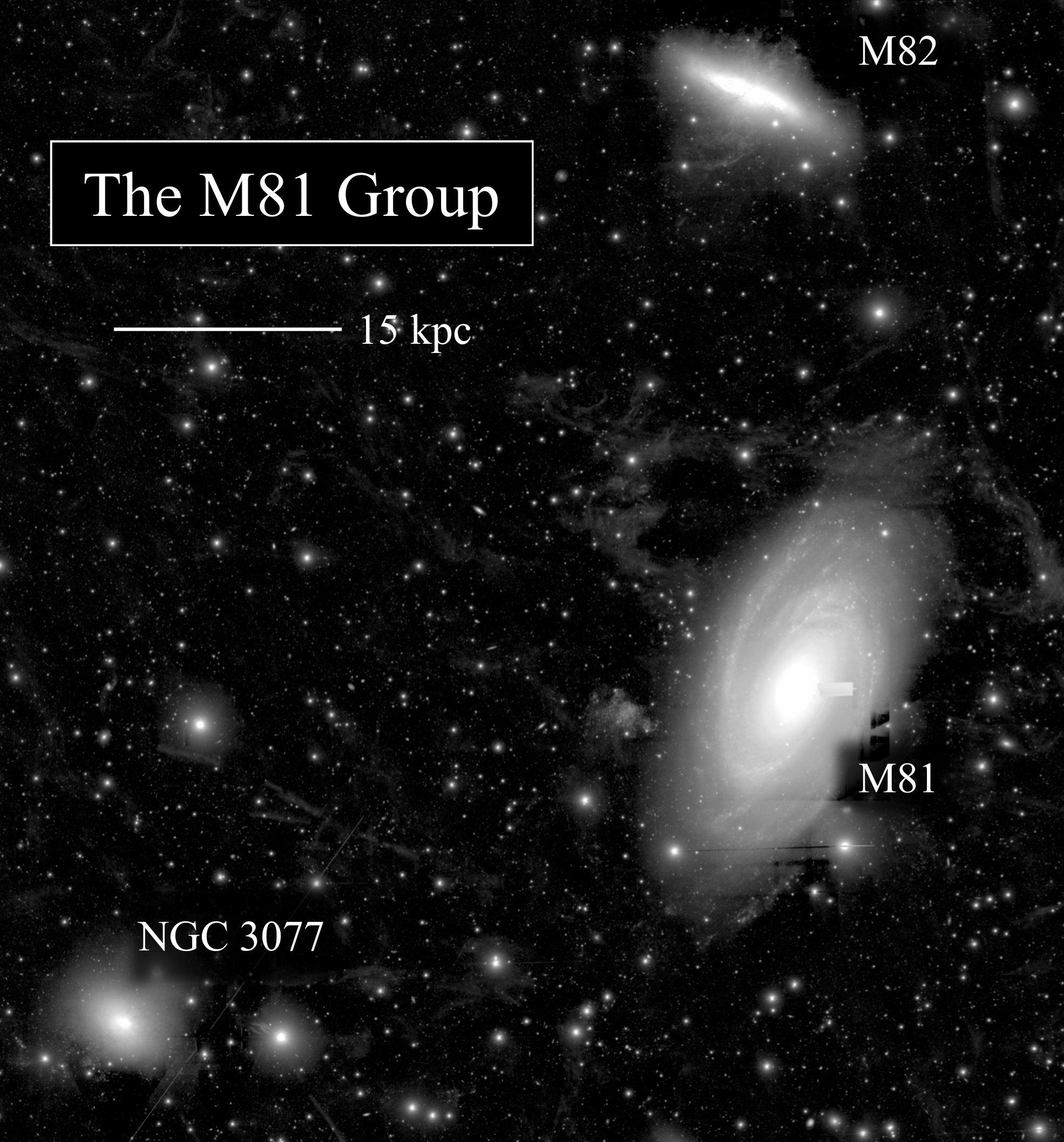} 
\caption{A deep, wide-field ($\sim$65\,kpc\,$\times$\,75\,kpc) $g$-band mosaic of the M81 Group, taken with Subaru HSC. A logarithmic stretch was used. The three primary interacting group members are labeled (M81, M82, and NGC 3077). The visible dark patches around the three galaxies, as well as bright stars, are the result of chip bleeds. The M81 Group is located behind a region of significant galactic cirrus, visible as patches of scattered light. This widespread cirrus impedes the inference of stellar halo properties through integrated light alone.} 
\label{fig:mosaic}
\end{figure*}

Fortunately, mergers also deposit a significant amount of loosely-bound stellar material which is retained within the DM halo --- the integral debris of all such events comprises the central galaxy's `stellar halo' \citep[e.g.,][]{spitzer&shapiro1972,bullock&johnston2005}. Stellar halos act as index fossils of past merger events, encoding the properties of these events long after their impact has been all-but-erased from typical observational diagnostics within the galaxy. Taking advantage of their close proximity, the stellar halos of the Milky Way (MW) and the Andromeda galaxy (M31) have been studied in exquisite detail, from their stellar populations \citep[e.g.,][]{bell2008,bell2010,ibata2014,gilbert2014,williams2015}, to their structure \citep[e.g.,][]{ibata2001,carollo2010,deason2011} and kinematics \citep[e.g.,][]{kafle2012,gilbert2018}. 

The stellar halos of a number of MW-mass galaxies in the Local Volume (LV) have also been studied in detail. As stellar halos of MW-mass galaxies are both large ($\sim$100\,kpc) and diffuse ($\mu_{V}\,{>}$\,28\,\magsqarc), there are several approaches which have been taken: (1) deep integrated light surveys \citep[e.g.,][]{merritt2016,watkins2016}, (2) deep `pencil beam' Hubble Space Telescope (HST) surveys which resolve individual stars \citep[e.g., GHOSTS;][]{radburn-smith2011,monachesi2016a,harmsen2017,jang2020}, and (3) wide field, ground-based surveys which resolve individual stars (e.g., M31, \citealt{ibata2014}; M81, \citealt{okamoto2015}; Cen A, \citealt{crnojevic2016}). Field-of-view, star--galaxy separation, and sensitivity to global properties are best optimized respectively with integrated light, resolved stellar populations with \textit{HST}, and ground based observations of resolved stars. Many nearby MW-like galaxies reside in regions of the sky plagued by significant galactic cirrus. This cirrus emission can substantially limit the sensitivity of integrated light to even bulk halo properties \citep[e.g.,][]{watkins2016,harmsen2017}. In these cases, resolved stellar populations are the optimal approach. 

These efforts have revealed that, among the $\sim$10 best-measured stellar halos of nearby MW-mass galaxies, there exists a spread of nearly \textit{two orders of magnitude in stellar halo mass}, and more than \textit{1\,dex in stellar halo metallicity} \citep[e.g.,][and references therein]{monachesi2016a,harmsen2017,bell2017}. Surprisingly, the MW and M31 sit on opposite ends of this distribution --- the MW being the least massive and metal-poorest \citep[e.g.,][]{bell2008}, while M31 is the most massive and metal-rich \citep[e.g.,][]{ibata2014} --- highlighting the enormous diversity in the accretion histories of MW-mass galaxies. 

Hints of this diversity in stellar halo properties have begun to appear in simulations \citep[e.g.,][]{monachesi2016b,dsouza&bell2018a,monachesi2019}, with indications that much of this diversity can be explained by the slope and scatter in the galaxy stellar mass--halo mass relation below $L^{*}$. Yet, the process of stellar halo assembly, and the associated mergers' impacts on the evolution of the central galaxies, is unclear. The question remains: \textit{how are these halos built?}
\begin{itemize}[topsep=3pt,noitemsep,leftmargin=12pt]
   \item It is now becoming clear from models that the most massive merger a galaxy experiences may dominate the observed properties of its stellar halo \citep[e.g.,][]{dsouza&bell2018a,dsouza&bell2018b,fattahi2019,lancaster2019}. Yet, what other important mergers did the galaxies experience before the largest event?
   \item Do large stellar halos require a higher number of substantial mergers over a galaxy's life, as seen in many simulations (e.g., \citealt{johnston2008,monachesi2019}), and interpreted from observations of galaxies such as M31 (e.g., \citealt{ibata2014,mcconnachie2018,mackey2019})? Or, can halo properties be dominated by a single merger? 
\end{itemize}

\noindent The mergers a galaxy experiences throughout its life are likely important drivers of its evolution. However, if stellar halo properties are, indeed, dominated by a single dominant merger, then the other substantial mergers a galaxy may have experienced will be effectively hidden from us for most systems. A powerful approach to address this observational impairment would be to study in detail the stellar halos of systems which are currently undergoing significant (i.e. dominant) mergers. This could simultaneously enable the inference of, and comparisons between, both their \textit{past} and \textit{future} largest mergers, and how such an event impacts the stellar halo. When combined with current measurements for non-merging systems, such an approach could shed invaluable light on the build-up of stellar halos and the evolution of MW-mass systems.

In this paper, we present a Subaru Hyper Suprime-Cam (HSC) survey of the resolved stellar halo populations of the interacting M81 Group (see Fig.\,\ref{fig:mosaic}; similar to the earlier survey of \citealt{okamoto2015}) --- the most detailed study of a stellar halo yet obtained outside of the Local Group (LG). The M81 Group is a quintessential example of a triple-interacting system --- hosting vast bridges of tidally stripped H\,I gas liberated from the two interacting satellites, M82 and NGC 3077 \citep{yun1994,deblok2018} --- and is the nearest ongoing significant merger (3.6\,Mpc; \citealt{radburn-smith2011}). Using a three-filter, equal-depth observing strategy, as well as numerous overlapping \textit{HST} calibration fields, we combine the relative advantages of `pencil beam' and ground-based surveys, revealing M81's stellar halo in never-before-seen detail. We use this new quantitative insight to show that, in a single merer event, M81 will span nearly the entire stellar halo mass--metallicity relation: transitioning from a low-mass, metal-poor halo, to one of the most massive, metal-rich halos known --- rivaled only by the halos of galaxies such as M31. 

\newcolumntype{s}{!{\extracolsep{25pt}}c!{\extracolsep{0pt}}}

\begin{deluxetable*}{cssss}
\tablecaption{\textnormal{Observations}\label{tab:obs}}
\tablecolumns{5}
\setlength{\extrarowheight}{-0.5pt}
\tablewidth{\linewidth}
\tabletypesize{\small}
\tablehead{%
\colhead{} & 
\multicolumn{2}{c}{Field 1} &
\multicolumn{2}{c}{Field 2} \vspace{1mm}\\ \cline{2-3} \cline{4-5}
\vspace{-2mm} \\
\colhead{} &
\colhead{} &
\colhead{Integration time$^b$} &
\colhead{} &
\colhead{Integration time} \\
\colhead{Filter} &
\colhead{\# Exposures$^a$} &
\colhead{(s)} &
\colhead{\# Exposures} &
\colhead{(s)} \\ \vspace{-3.5mm}
}
\startdata
$g$ & 14 & 4200 & 18 & 5400 \\
$r$ & 11 & 3300 & 12 & 3600 \\
$i$ & 11 & 3300 & 11 & 3300 \\
\enddata
\tablecomments{$^a$\,Total number of 300\,s exposures for a single field. $^b$\,Total integration time (i.e. 300\,s\,$\times$\,$N_{exp}$).}
\end{deluxetable*}

\section{Observations}
\label{sec:reduction}
These observations were taken with the Subaru HSC, through the Gemini--Subaru exchange program (PI: Bell, 2015A-0281). Imaging was undertaken in the `classical' observing mode over the nights of March\,26--27, 2015. The survey consists of two pointings (each $\sim$1\fdg5 FOV), in each of three ($g,r,i$) filters. Pointings were primarily chosen to fully cover the outer regions of all three interacting galaxies --- M81, M82, and NGC 3077. Integration times for each field+filter combination are given in Table\,\ref{tab:obs}. Differences in observing time between the two fields in the same filter reflect adjustments made in response to changing conditions (e.g., sky transparency, background, and seeing).

The data were reduced with the HSC optical imaging pipeline \citep{bosch2018}. The pipeline performs photometric and astrometric calibration using the Pan-STARRS1 catalog \citep{magnier2013}, but reports the final magnitudes in the HSC natural system, which we then finally correct to the SDSS filter system. The version of the pipeline adopted here performs background subtraction with an aggressive 32-pixel mesh, optimizing point-source detection and removing most diffuse light. Sources are detected in all three-bands, though $i$-band is prioritized to determine reference positions for forced photometry. Forced photometry is then performed on sources in the $gri$\ co-added image stack.

All magnitudes were corrected for galactic extinction following \cite{schlafly&finkbeiner2011}. We find that, broadly, the M81 Group has relatively consistent E(B--V)\,$\simeq$\,0.1. However, the innermost regions of M82 suffer `contamination' from dust emission, causing artificially higher estimated extinction. Because of this, we limit E(B--V) to a maximum of 0.1 in the region of M82. Image depth was nearly uniform across the two fields, yielding extinction-corrected point source detection limits of $g\,{=}\,27$, $r\,{=}\,26.5$, and $i\,{=}\,26.2$, measured at $\sim$5$\sigma$. See \cite{bosch2018} for an in-depth discussion of the photometric uncertainties output by the HSC pipeline. Seeing was relatively stable, resulting in consistent point-sources sizes of 0\farcs7--0\farcs8 down to the detection limits. 

\section{Star--Galaxy Separation \& RGB Selection}
\label{sec:stargal}

For galaxies such as M81, which are well beyond the Local Group ($D_{\rm M81}\,{\simeq}$\,3.6 Mpc; \citealt{radburn-smith2011}), the bulk of the resolvable stellar populations (i.e. the stellar main sequence) is too faint to observe. In M81, for example, the main-sequence turn-off of the average halo population (e.g., Age\,$\sim$\,9\,Gyr, [M/H]\,${\sim}\,{-}1.2$; \citealt{durrell2010}) occurs around $i\,{\sim}\,31$. Characterization of the stellar halo populations therefore requires a more luminous sub-population to trace the underlying stellar population. Red giant branch (RGB) stars are numerous, luminous, and are well-tied to the underlying stellar population, making them excellent tracers. We detect RGB stars to two magnitudes below the tip (the `TRGB').

At the depths achieved by this survey (i.e. $g\,{\lesssim}\,27.8$, $r\,{\lesssim}\,26.8$, $i\,{\lesssim}\,26.2$), the majority of detected sources are background galaxies, rather than stars in M81's halo. As an example, an initial morphological cut selecting sources with FWHM\,$\leqslant$\,0\farcs75 eliminates \textit{80\% of sources from our catalog}. For shallower ground-based observations \citep[e.g., the PAndAS survey][]{ibata2014}, detected background galaxies at the relevant magnitudes are typically more morphologically distinct than at deeper limits, and such a cut results in reasonable star--galaxy separation. Likewise, for \textit{HST} observations, despite reaching comparable limits to this survey, the majority of even faint high-redshift galaxies are morphologically distinguishable from stars \citep[see e.g.,][]{radburn-smith2011}. 

\begin{figure*}[!ht]
\leavevmode
\centering
\includegraphics[width={0.83\linewidth}]{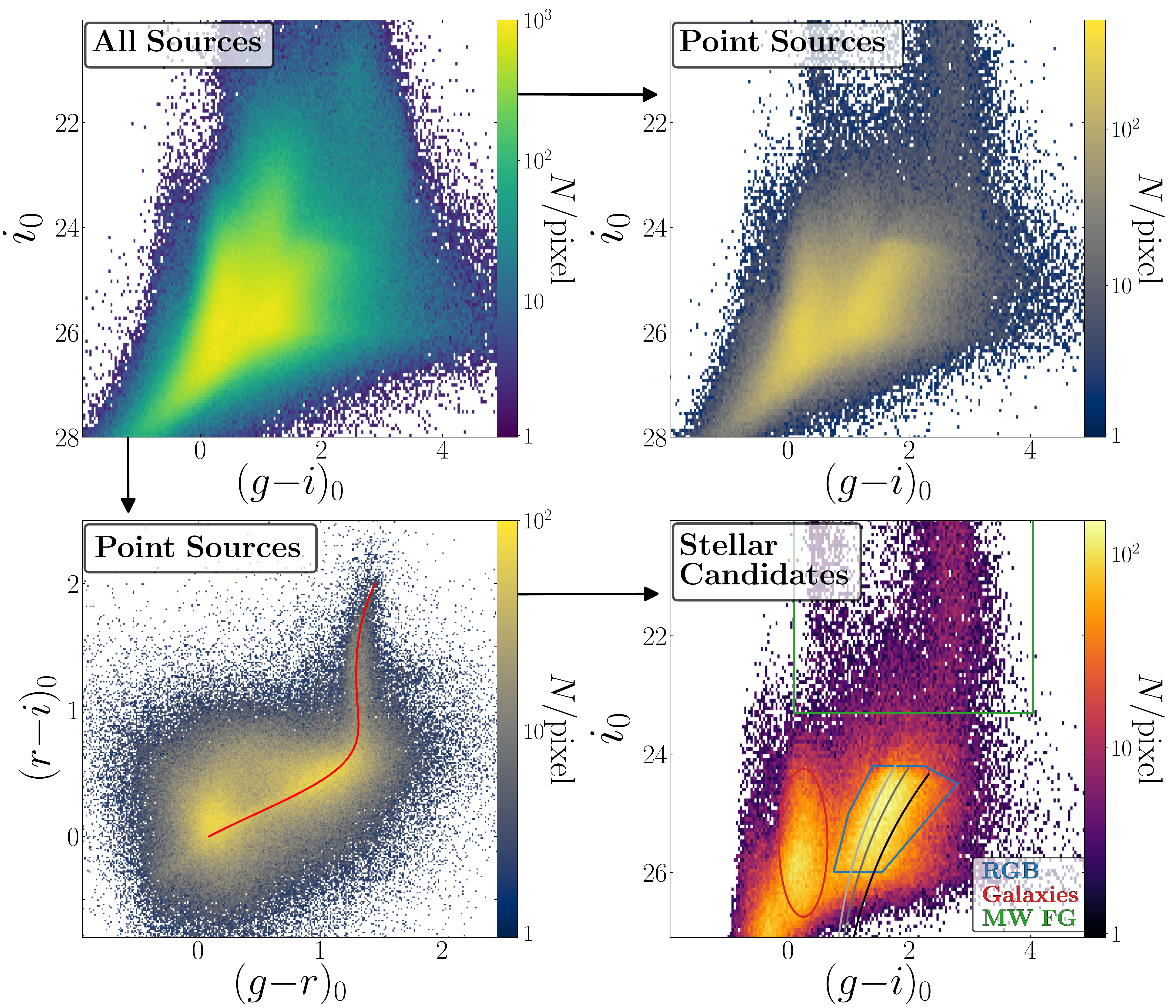}
\caption{\textbf{\uline{Top left}:} $g{-}i$\ vs. $i$\ CMD of all detected sources in our survey footprint. \textbf{\uline{Top right}:} $g{-}i$\ vs. $i$\ CMD of point sources, with size $\leqslant$0\farcs75. \textbf{\uline{Bottom left}:} Color--color diagram of point sources, with size $\leqslant$0\farcs75. The stellar locus is shown as a red curve. Only sources lying on the stellar locus, within $\sigma$+0.2\,mag, are selected as stellar candidates. \textbf{\uline{Bottom right}:} $g{-}i$\ vs. $i$\ CMD of all morphologically ($<$0\farcs75) and color-selected (${<}\sigma$+0.2\,mag from SL) stellar candidate sources. The locus of unresolved background galaxies (red ellipse) is now easily distinguishable from the RGB selection box (blue). The region of the CMD dominated by MW foreground stars is shown in green. Three stellar isochrone models \citep[PARSEC;][]{bressan2012} are shown (age = 10\,Gyr), with metallicities of [Fe/H] = $-$2, $-$1.5, and $-$1.}
\label{fig:cmd}
\end{figure*}

\begin{deluxetable*}{sss}
\floattable
\tablewidth{0.8\textwidth}
\tablecaption{\textnormal{RGB Selection Criteria}\label{tab:select}}
\tablecolumns{3}
\setlength{\extrarowheight}{-0.5pt}
\tablewidth{\linewidth}
\tabletypesize{\small}
\tablehead{%
\colhead{Type} &
\colhead{Description} &
\colhead{Criterion} \\ \vspace{-3.5mm}
}
\startdata
\vspace{-2mm} \\
 & & $\theta_{x}(g)\,{<}$\,0\farcs75~~~$\theta_{y}(g)\,{<}$\,0\farcs75 \\
\textbf{Morphological} & Size constraints in each filter and along each axis & $\theta_{x}(r)\,{<}$\,0\farcs75~~~$\theta_{y}(r)\,{<}$\,0\farcs75 \\
 & & $\theta_{x}(i)\,{<}$\,0\farcs75~~~$\theta_{y}(i)\,{<}$\,0\farcs75 \vspace{1mm} \\
\textbf{Color--Color} & Proximity to stellar locus in $g{-}r$\ color & $|(g{-}r)-(g{-}r)_{\rm SL}|\,{<}\,\sigma_{g{-}r}$\,+\,0.2 \vspace{1mm}\\
 & & ($g{-}i$,~$i$) = \\
\textbf{Color--Magnitude} & Vertices of the $g{-}i$\ vs. $i$\ RGB selection box & (0.75,26.0), (1.55,26.0), (2.8,24.5), \\
 & & (2.25,24.2), (1.4, 24.2), (1.0,25.0) \\
\enddata
\tablecomments{\uline{Morphological}: Size is FWHM along each axis. \uline{Color--Color}: $(g{-}r)_{\rm SL}$\ is the $g{-}r$\ color of the stellar locus at a given $r{-}i$. $\sigma_{g{-}r}$\ is the measured source uncertainty in $g{-}r$.}
\end{deluxetable*}

It is at the interface reached by this survey --- deep detection limits, yet ground-based image quality --- where star--galaxy separation becomes truly challenging. In this regime, many faint background galaxies are as equally point-like as stars, motivating selection criteria beyond morphological cuts. As they are amalgams of numerous stellar populations, galaxies exist at virtually every position in the color--magnitude diagram. Many distant galaxies are located at relatively bluer $g{-}i$\ colors compared to RGB stars, resulting in a CMD feature located at $g{-}i\,{\sim}\,0.1$. However, selecting RGB stars by their position in the CMD does not eliminate contamination from background galaxies. 

Fortunately, stars inhabit an empirical `stellar locus' (SL) in broadband (e.g., $g{-}r$/$r{-}i$) color--color space \citep[e.g.,][]{covey2007,ivezic2007,high2009,davenport2014}. Our addition of the $r$\ filter allows us to leverage this distinct color--color information to distill our RGB sample by an additional 30\%. `Stars' are classified as sources $<$0\farcs75 in size (along both axes) and with $g{-}r$\ distance from the SL $< \sigma_{g{-}r}$\,+\,0.2\,mag at a measured $r{-}i$\ color, where $\sigma_{g{-}r}$\ is the $g{-}r$\ photometric color uncertainty and 0.2\,mag is the adopted systematic width of the SL (from \citealt{covey2007,high2009}; see also \citealt{smercina2017}). Figure\,\ref{fig:cmd} demonstrates this selection process, showing the CMD and color--color diagrams of \textit{all} sources, as well as the final, distilled CMD following our selection algorithm. Though the RGB is easily distinguishable using the SL, the unresolved background galaxy locus at blue colors remains. The locations of each are marked. This choice reflects the required sensitivity for faint SB and color measurements, which prioritize the purity of the RGB sample, rather than overall completeness. Table\,\ref{tab:select} gives the parameters for our selection process. The resulting culled sample of 73,946 RGB stars is used throughout the rest of the paper.

\begin{figure*}[t]
\leavevmode
\centering
\includegraphics[width={0.81\linewidth}]{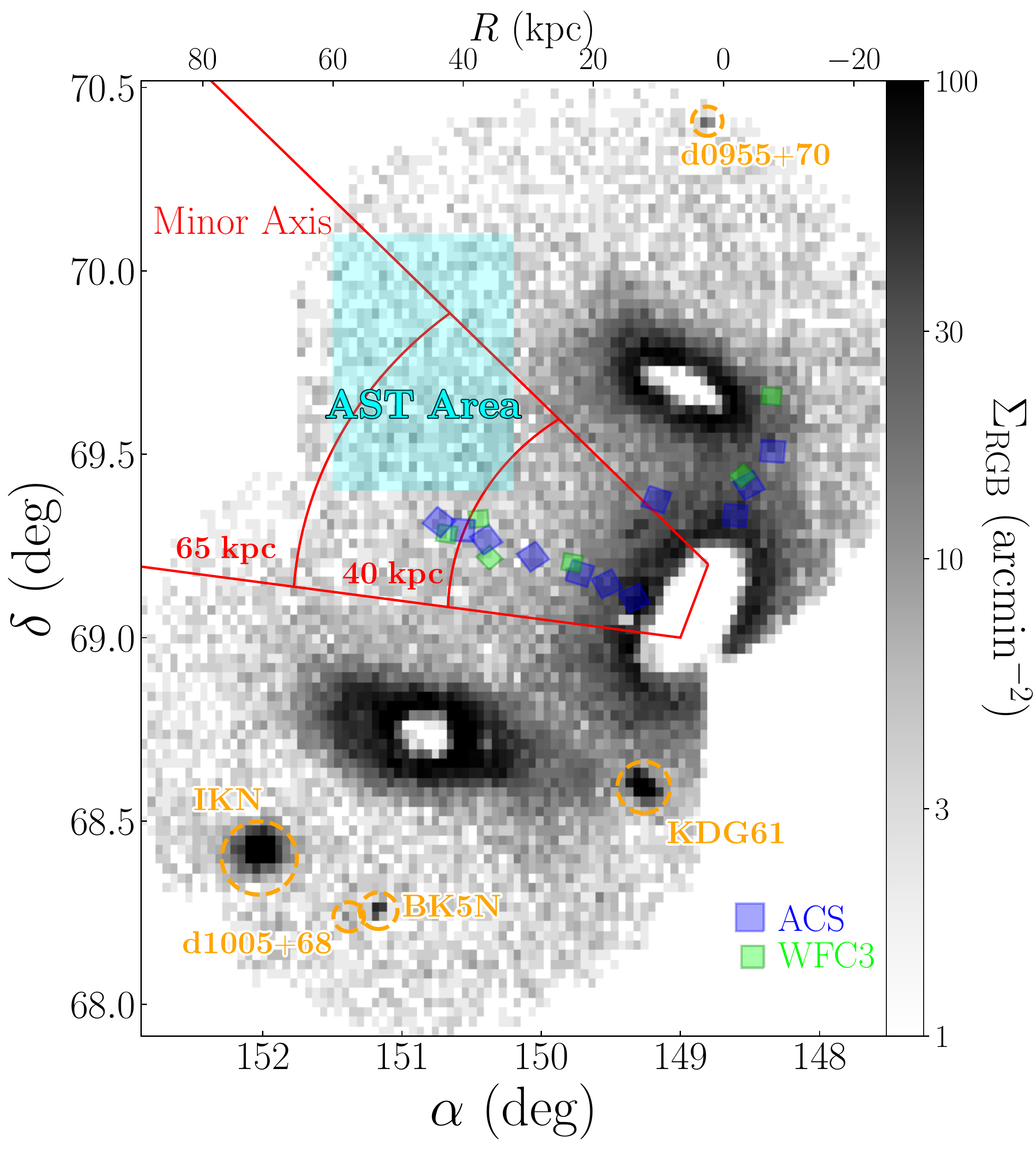}
\caption{\textbf{\uline{Left}:} Grayscale logarithmic density map of RGB stars in M81's halo. This map shows observed RGB candidate counts (defined as in \S\,\ref{sec:stargal}), witout correction. A colorbar showing the mapping to density is given on the right. Existing \textit{HST} fields from the GHOSTS survey \citep[e.g.,][]{radburn-smith2011,monachesi2013} are overlaid (\textcolor{blue}{ACS}---blue/\textcolor{green}{WFC3}---green). The region defined as M81's `minor axis' in this paper is shown in red, along with the 40\,kpc and 65\,kpc radii used for measuring accreted mass and definining SB limti, respectively. Additionally, the spatial region in which our suite ASTs were peformed (\S\,\ref{sec:ast}) is shown as a cyan box. Lastly, all low-mass satellites of M81 within the survey area are circled and labeled in orange, for the reader's reference.}
\label{fig:field-overview}
\end{figure*}

\begin{figure*}[t]
\leavevmode
\centering
\includegraphics[width={\linewidth}]{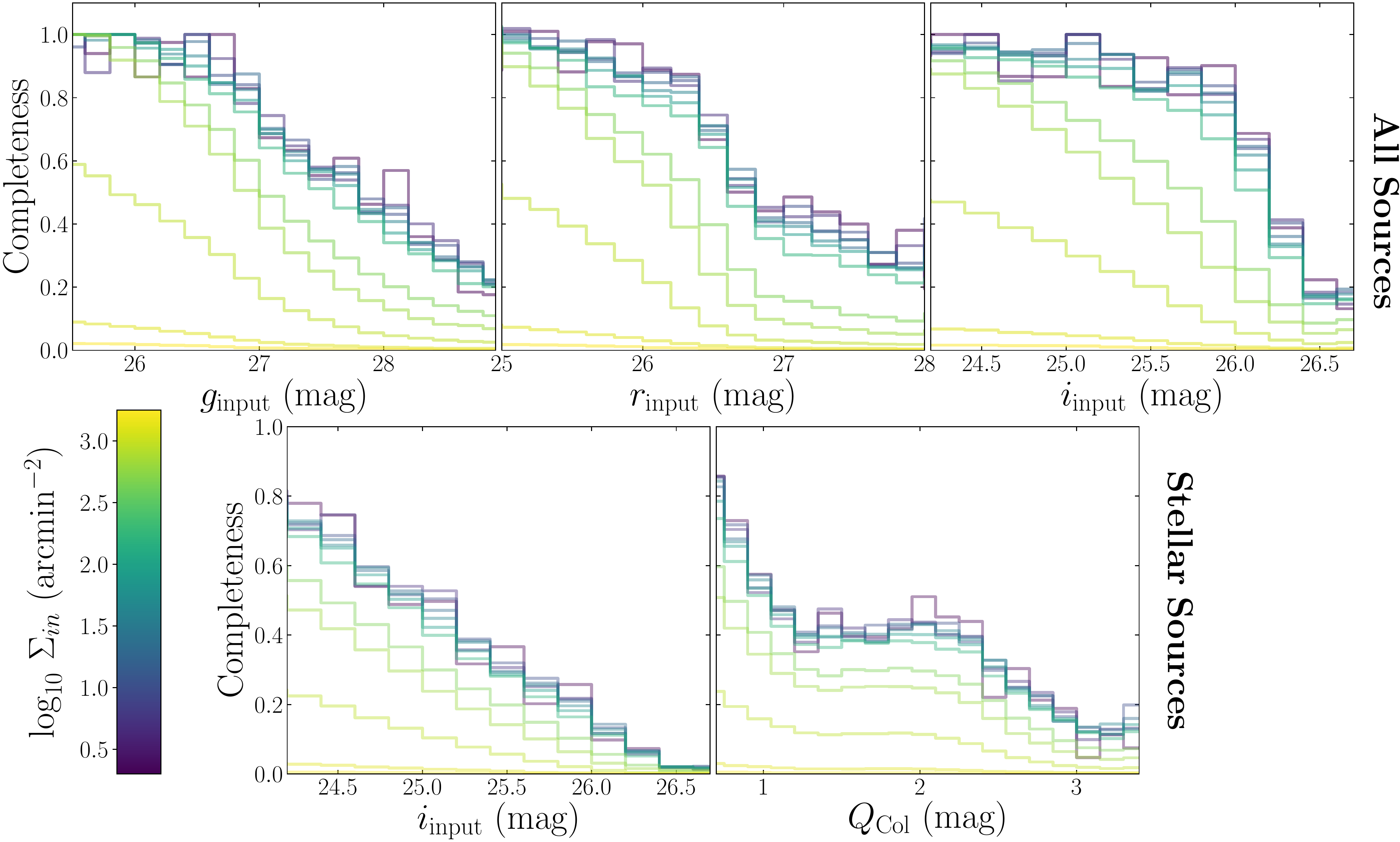}
\caption{Photometric completeness functions derived from our twelve ASTs. `Completeness' is defined as the fraction of sources in the initial AST catalog with an observed positional match within $<$0\farcs3. The initial input density of each AST is color-coded according to the colorbar given at the bottom left. \uline{Top row}: Completeness curves are shown for \textit{all sources} as a function of input magnitude in each filter: $g$-band (left), $r$-band (center), and $i$-band (right). \uline{Bottom row}: Completeness as a function of $i$-band magnitude (left) and $Q$-color (right) for stellar candidate sources (selection described in \S\,\ref{sec:stargal}; Table \ref{tab:select}). See \S\,\ref{sec:ghosts-cal} for an explanation of the transformation between $g{-}i$\ and $Q_{\rm Col}$.}
\label{fig:ast-complete}
\end{figure*}

\section{Calibration of Stellar Populations}
\label{sec:calibration}
Though our sample of RGB stars is highly pure, due in large part to the addition of the $r$-band filter and excellent ground-based image quality, we face a number of competing issues which work to inhibit quantitative inferences from the observed stellar populations --- mainly: (1) remaining contamination (from background galaxies), (2) crowding, and (3) incompleteness. These effects impact our ability to accurately measure the properties of the observed stellar populations, particularly their colors and average density on the sky. To account for and quantify these effects, we conduct a series of a artificial star tests (ASTs), as well as compare against existing \textit{Hubble Space Telescope (HST)} observations from the GHOSTS survey (similar to the strategy adopted by \citealt{bailin2011} for NGC 253). In this section, we first describe the AST experiment design, followed by an overview of the existing GHOSTS observations, and finally describe the derived corrections we adopt, to both observed densities and colors (to be used in \S\,\ref{sec:results}). Figure \ref{fig:field-overview} gives an overview of the survey area, including positions of the area in which ASTs were performed, the existing GHOSTS fields, and the minor axis region defined in this paper (for use in \S\,\ref{sec:results}), all shown on the uncorrected density map of RGB stellar candidates.

\subsection{Artificial Star Tests}
\label{sec:ast}
In an effort to quantify photometric completeness in our survey, and the impact on our RGB-based metrics, we generated a suite of twelve artificial test experiments, successively increasing the injected source density logarithmically, from 2 stars/arcmin$^{2}$\ for the least dense experiment to 1,810 stars/arcmin$^{2}$\ for the most dense. A catalog of 100,000 artificial stellar sources was created with $g{-}i$\ colors drawn uniformly between 0.5--3\ and $i$-band magnitudes drawn uniformly between 24--26.7. $r$-band magnitudes were then calculated assuming each artificial star lies on the $g{-}r$/$r{-}i$\ stellar locus (see \S\,\ref{sec:stargal}). We then drew uniformly from this catalog and injected sources with PSF equivalent to the average seeing across the field (\S\,\ref{sec:reduction}), at the twelve density levels, within a 0.32 deg$^{2}$\ region intrinsically sparse in RGB stars (to the northeast of M81; see Figure \ref{fig:field-overview}). This is the first implementation of ASTs processed through the Subaru HSC pipeline. As the machinery for processing these ASTs is still nascent, it was simpler to perform the ASTs in an uncrowded region of the survey footprint and increase the density manually, rather than the more standard approach of injecting artificial sources across the entire footprint. As such, our assessment of photometric quality is only directly applicable to this region. The existing \textit{HST} data is fundamental to assessing the generalizability of our AST results to the rest of the survey footprint.

Following injection of the artificial sources into the raw images, we processed each AST using the HSC pipeline, in the same manner as the observations (see \S\,\ref{sec:reduction}) --- producing full pipeline photometric catalogs for each. The observed catalog for each experiment were then matched against the input AST catalog, with a recovered match defined as the closest detected source within 0\farcs3 of the input source position. In the lowest-density case, our 0\farcs3 matching criterion yields an 83\% global completeness amongst nearest neighbors matches to the input AST catalog.

In Figure \ref{fig:ast-complete} we show the results of the AST experiments, showing completeness functions for all sources in each filter, as well as $i$-band and color completeness functions for stellar-selected sources. We find that for input source densities $\lesssim$\,100--150 stars arcmin$^{-2}$, photometric completeness is independent of input density, but begins to decrease sharply at densities $\gg$100 stars arcmin$^{-2}$. We also note the difference in the slope of the completeness function when considering all sources, compared to stellar candidates. At fainter magnitudes, the greater precision in both morphology and photometry required by our stellar selection criteria results in a much steeper completeness function than for the full sample. The three highest-density ASTs show a dramatic decrease in completeness, with almost no stellar candidates detected, indicating that nearly all sources detected by the pipeline are photometric blends. These densities are consistent with the densities seen in M81's disk, high enough to be robustly detected as integrated light, and thus are excluded, rather than corrected for, in our resolved stellar halo analysis.

\begin{figure}[t]
\leavevmode
\centering
\includegraphics[width={\linewidth}]{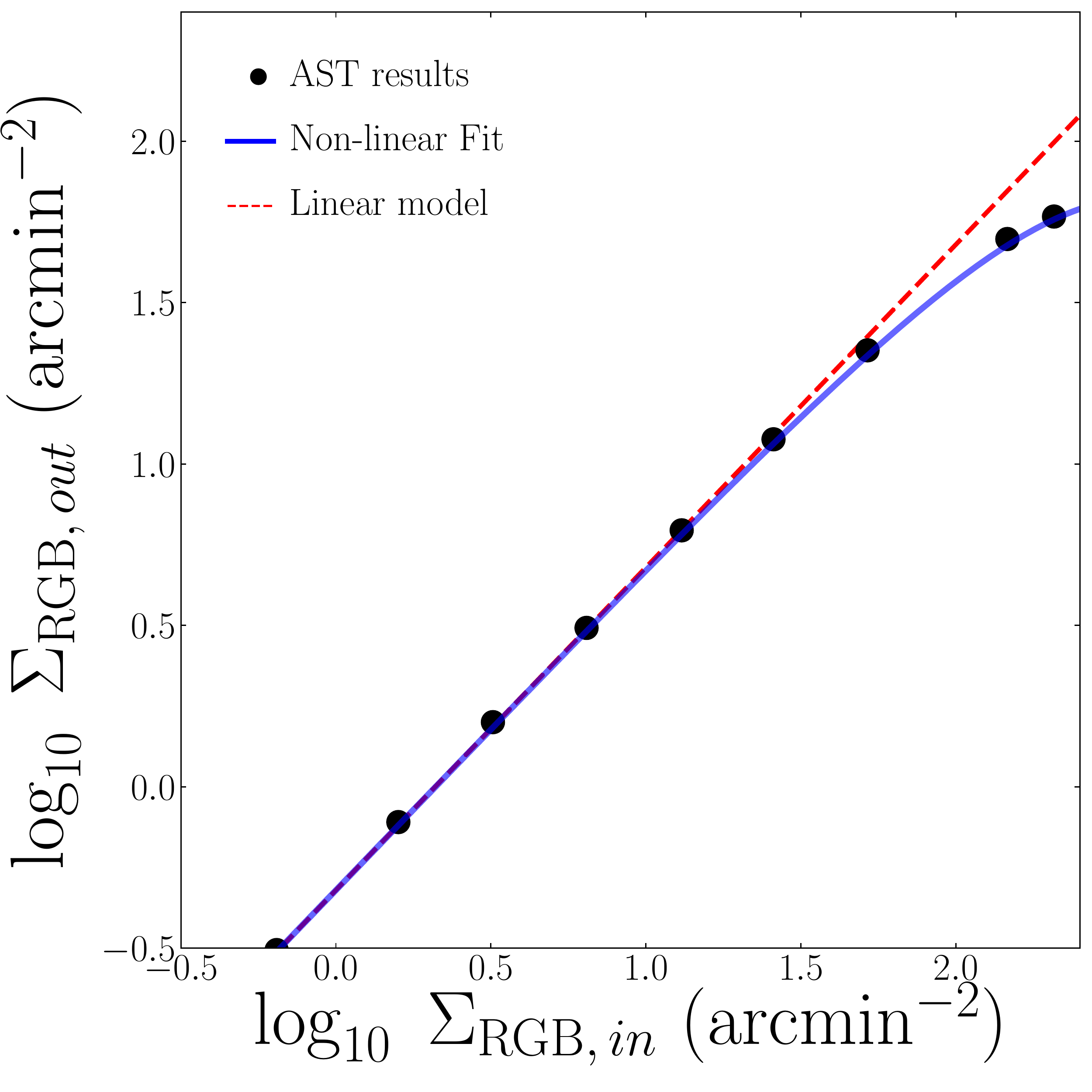}
\caption{Observed correspondence (in arcmin$^{-2}$)\ between the density of RGB-selected stars in each AST catalog and the recovered RGB candidates in each observed catalog. A best-fit linear model is shown in red, which the results follow for densities ${\sim}10$\ RGB stars/arcmin$^2$. Additionally, we show in blue the best-fit non-linear model given in \S\,\ref{sec:denscal}.}
\label{fig:AST-RGB}
\end{figure}

\subsubsection{RGB Density Calibration}
\label{sec:denscal}
Given the crowding- and magnitude-dependent impacts on photometric completeness shown in Figure \ref{fig:ast-complete}, any inferences made from RGB star counts in a given region of sky, selected as described in \S\,\ref{sec:stargal}, must account for these effects. To calibrate RGB counts specifically, we apply the stellar selection criteria, and RGB selection box (given in Table \ref{tab:select}), to both the input AST catalog and the recovered catalog to assess the correspondence between input and recovered RGB density, as a function of true source density. 

Figure \ref{fig:AST-RGB} shows the results of this analysis, with average input RGB density (in arcmin$^{-2}$) across the AST area plotted against the average recovered RGB density, for each of the twelve ASTs. As expected from the completeness functions in Figure \ref{fig:ast-complete}, the lower density half of the ASTs (with $\lesssim$\,100\,stars arcmin$^{-2}$\ initial density) exhibit a smooth, linear correspondence between input and recovered RGB counts. The normalization factor is 0.48 --- i.e. approximately 48\% of input RGB-like stars are recovered in the linear density regime. However, for large input densities ($\gg$100\,stars arcmin$^{-2}$), the correspondence between input and recovered counts deviates non-linearly, presumably due to the increasing impact of crowding. Though we take a somewhat non-standard approach to AST analysis, the reliable behavior of the AST results relative to expectations, as well as the consistency with empirical comparisons to existing \textit{HST} observations, suggests that this is an effective approach. We fit the results with a model of the form
\begin{equation}
\label{eq:1}
    \Sigma_{\rm out} = \Sigma^*\,\Sigma_{\rm in}\,e^{-\alpha\,\Sigma_{\rm in}/\Sigma_0}, 
\end{equation}
with best-fit parameters of $\Sigma^*\,{=}\,0.48$, $\Sigma_0\,{=}\,750$, and $\alpha\,{=}\,2$. In \S\,\ref{sec:sb-prof}, we will use these results to correct observed RGB density to the intrinsic density in order to measure SB and stellar mass.

\begin{figure}[t]
\leavevmode
\centering
\includegraphics[width={\linewidth}]{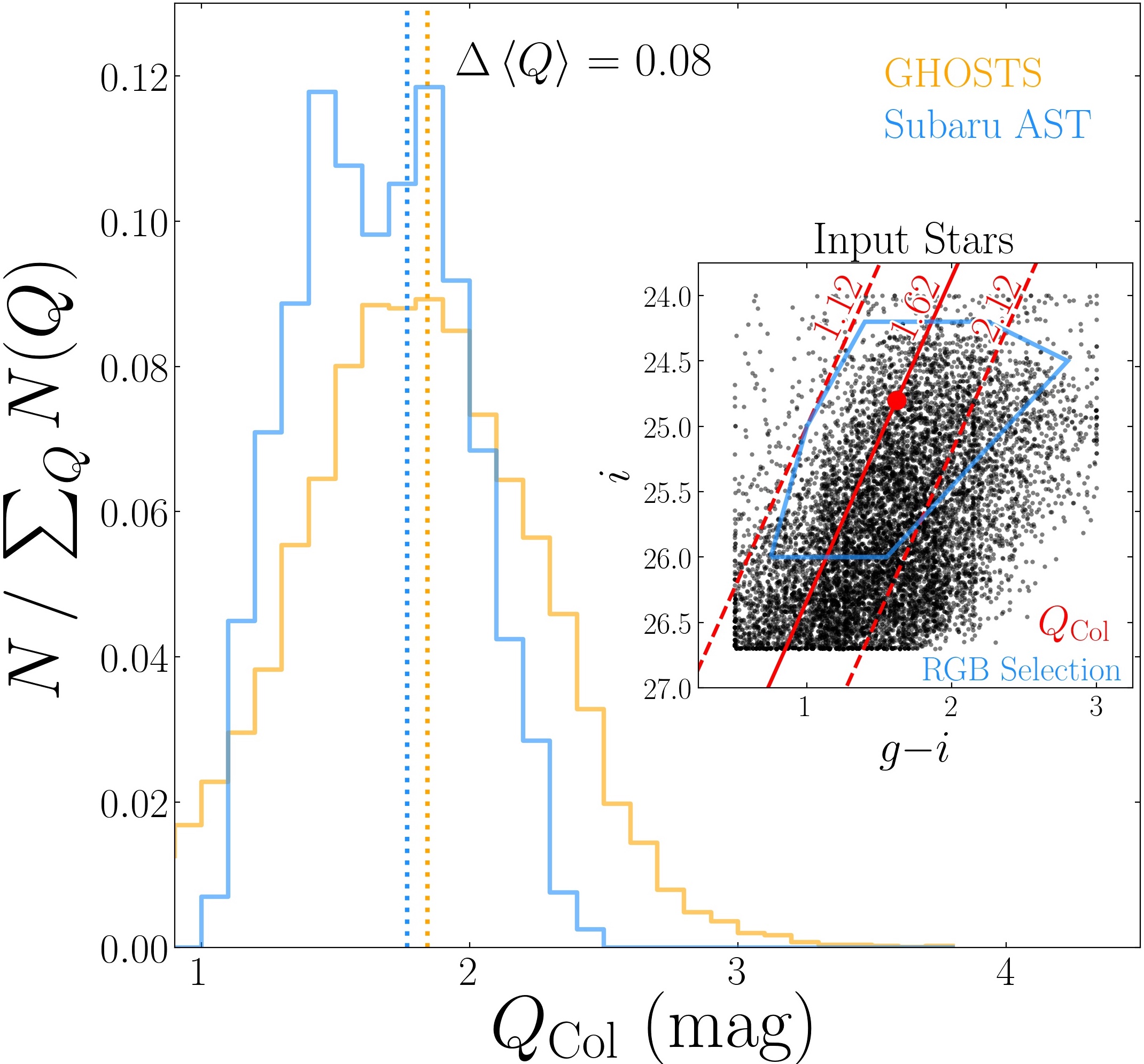}
\caption{Comparison of the normalized $Q$-color distributions for RGB stars in the GHOSTS survey fields (orange), and observed RGB candidates in the HSC catalog, recovered from an input CMD of GHOSTS stars matched to stars in one of our AST catalogs (blue). The median of each distribution is shown as a dashed line. \uline{Right inset}: CMD of the input GHOSTS catalog used to construct the $Q_{\rm Col}$\ distributions, converted to $g$\ \& $i$\ filters and matched to the AST catalog. The RGB selection box is shown in blue and curves of three constant $Q$-color are overlaid: $Q_{\rm Col}$\,=\,1.12, 1.62, and 2.12. The rotation point (1.62,24.8) at which we define $Q_{\rm Col}$\ is shown as a red dot. The substantial population of red sources in the GHOSTS fields that exist outside of the Subaru RGB selection box in color--magnitude space results in a significant offset in median color of 0.08\,mag.}
\label{fig:ghosts-qcol}
\end{figure}

\subsubsection{Comparisons to GHOSTS}
\label{sec:ghosts-cal}
Perhaps the more nuanced measurement of the observed stellar populations is that of color, and in turn estimates of abundance. Our survey is optimally geared to efficiently detecting RGB stars at colors of $g{-}i$\,=\,1--1.5. For M81, this corresponds to limiting $g$-band magnitudes of $\sim$27. However, the most metal-rich RGB stars, i.e. those with [M/H]\,${\gg}\,{-}0.5$, will have $g$-band magnitudes of 28--29 --- substantially fainter than the depths achieved by  this survey. Therefore, unless $g$-band observations are substantially deeper than $i$-band, any metal-rich populations that might exist will be too faint to observe in this survey, and all similarly-designed ground-based surveys. 

However within our HSC footprint, there are 17 existing \textit{HST} fields (ACS and WFC3) with high-quality stellar catalogs from the GHOSTS survey \citep{radburn-smith2011,monachesi2013}. These fields span the majority of the stellar density scale probed by our ASTs, ranging from $\sim$5--400 stars arcmin$^{-2}$. A number of these observed fields have been used to robustly measure the minor axis color \citep{monachesi2016a} and SB \citep{harmsen2017} profiles of nearby galaxies, including M81, using RGB stars detected in the F606W/F814W filters. While covering far less of the M81 Group than our HSC observations, the GHOSTS fields are $\sim$1\,mag deeper in $i$-band and can effectively probe the redder stellar populations unobserved by HSC. We show in \S\,\ref{sec:sb-prof} that this difference in photometric depth does not impact our SB analysis. However, in the event that there is a substantial subset of higher-metallicity stellar populations, the shallower color sensitivity of HSC could impact our ability to accurately measure the color, and therefore metallicity, of the intrinsic halo populations. 

To assess the presence of such populations, and the impact on measured color they might effect, we use the existing GHOSTS fields in combination with our AST catalogs. We take the composite CMD of 14 of the GHOSTS fields and we convert from F606W$-$F814W vs. F814W to $g{-}i$\ vs. $i$. Following \cite{monachesi2013} and \cite{harmsen2017}, we use a [M/H]\,=\,$-$1.2, 10\,Gyr old PARSEC \citep{bressan2012} isochrone model for the conversion, though we note that the $g{-}i$/F606W$-$F814W color--color relation for RGB stars is relatively insensitive to metallicity for halo-like populations. Using this $g{-}i$/$i$\ converted catalog of GHOSTS sources, we matched each to the nearest star in the CMD of one of our AST catalogs (of intermediate density; 80\,stars arcmin$^{-2}$). This input star catalog was then matched to the corresponding pipeline-produced source catalog and RGB star candidates were selected following \S\,\ref{sec:stargal}. The resulting CMD of GHOSTS--AST matched sources is shown in Figure \ref{fig:ghosts-qcol} (inset right). 

\cite{monachesi2016a} measured color profiles along the major and minor axes of the GHOSTS sample, including M81. To measure more robust colors, which also are intrinsically better-tied to population changes due to metallicity, they adopt a revised color metric, $Q$\ \citep[see also][]{monachesi2013}. As the isochrone model curve for a metal-poor stellar population is nearly a straight line for the upper portion of the RGB, the $Q$-color corresponds to a CMD which has been rotated around a point 0.5\,magnitudes below the TRGB, such that the RGB is nearly vertical. We adopt the $Q$\ metric for this paper as well, for all of our color-based analysis. As we are operating in the $g{-}i$\ filters, we define a new $Q_{\rm Col}$\ corresponding to a rotation angle of $-$22\textdegree. An example of this redefined color schema is shown overlayed on the CMD of GHOSTS--AST matched input stars in Figure \ref{fig:ghosts-qcol} (inset right). 

The $Q_{\rm Col}$\ distributions of both the converted GHOSTS stellar catalog and the sources recovered from the GHOSTS stars matched to the input AST catalog are shown in Figure \ref{fig:ghosts-qcol}. The $Q_{\rm Col}$\ distribution of HSC-observed RGB star candidates shows a distinct deficit of red sources, relative to GHOSTS, amounting to an offset of 0.08\,mag bluewards in the median $Q_{\rm Col}$. This offset represents a difference in the median measured and median intrinsic colors of the halo populations. We thus take this offset into account when reporting median color measurements, and related median metallicities, throughout the rest of the paper.

\begin{figure*}[t]
\leavevmode
\centering
\includegraphics[width={\linewidth}]{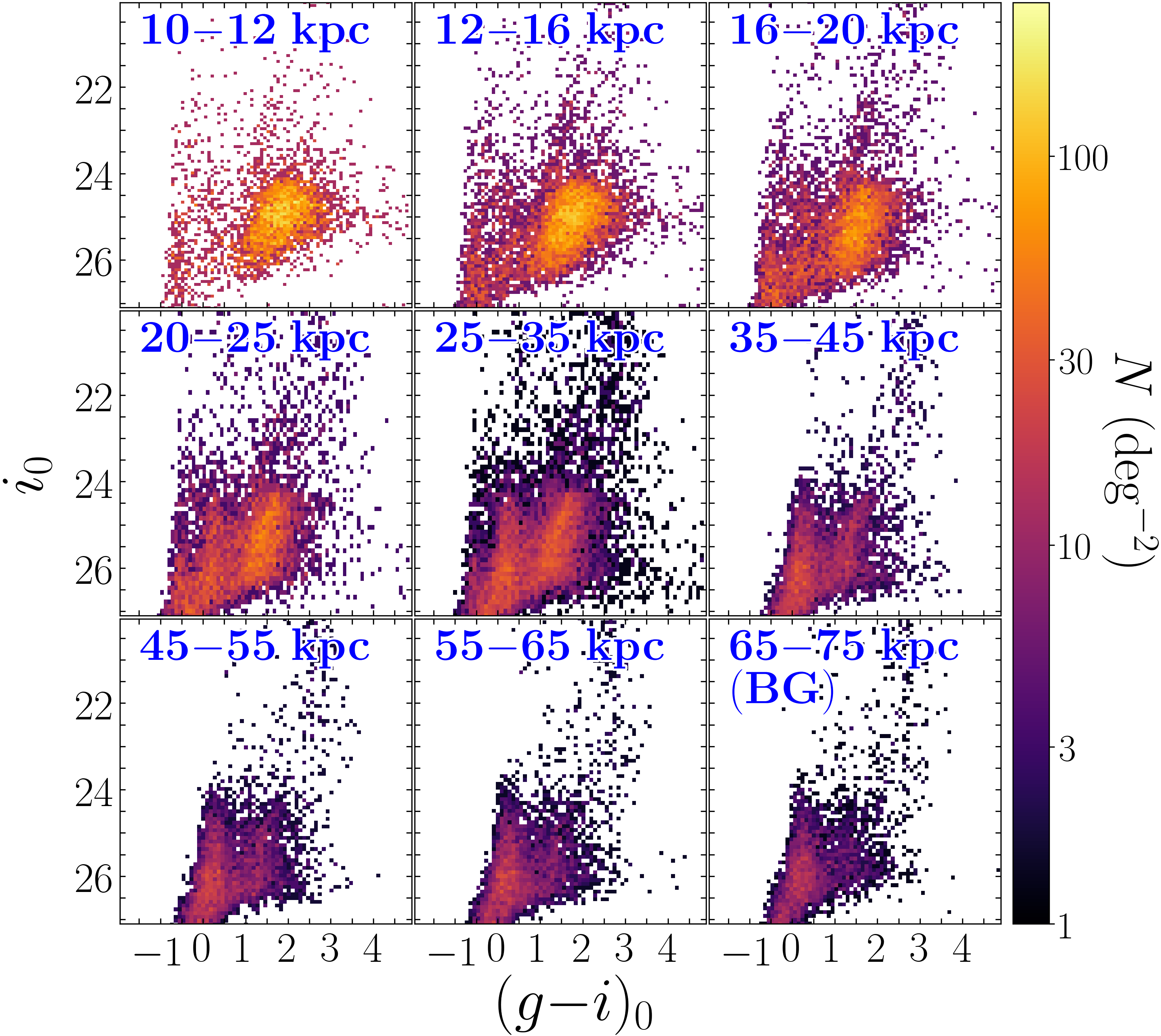}
\caption{$g{-}i$\ vs.\ $i$\ CMDs of stellar-selected sources along M81's minor axis, displayed in 9 radial bins out to 75\,kpc. The radial range for each CMD is given in blue. CMDs are displayed as Hess diagrams (i.e. density bins), with the counts scaled to density per deg$^2$\ in the given radial section. While the RGB is prominent at most radii, it decreases in strength with increasing radius, indicating a steep negative density profile. By the 65--75\,kpc radial bin RGB-colored sources are substantially less numerous, and a clear RGB is not visible; the CMD is dominated by background galaxies and MW foreground stars.}
\label{fig:minax-cmds}
\end{figure*}

\begin{figure}[t]
\leavevmode
\centering
\includegraphics[width={\linewidth}]{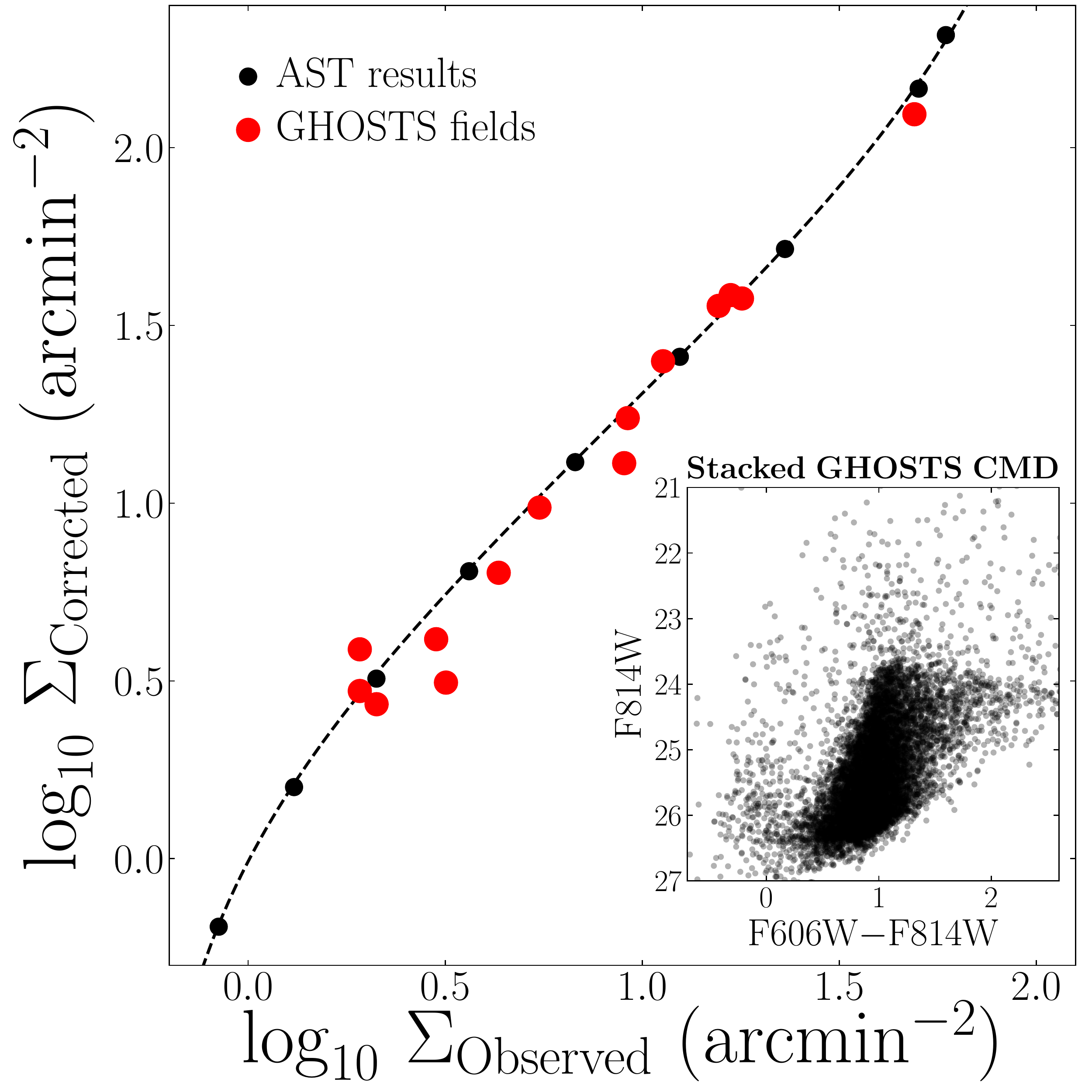}
\caption{Observed RGB density vs. the `corrected' density, accounting for background subtraction, completeness, and crowding. The black points show the relationship derived from each AST, where the corrected density is equivalent to the input RGB density from the pure AST catalogs. The best-fit curve is given as dashed line, with functional form following Equation \ref{eq:2}. The red points show the RGB density observed with Subaru in each of the GHOSTS fields plotted against the RGB densities published by \cite{harmsen2017}. The stacked \textit{HST} F606W$-$F814W CMD of all 14 GHOSTS fields is shown for reference at the bottom right.}
\label{fig:ast-ghosts-density}
\end{figure}

\begin{figure*}[t]
\leavevmode
\centering
\includegraphics[width={0.8\linewidth}]{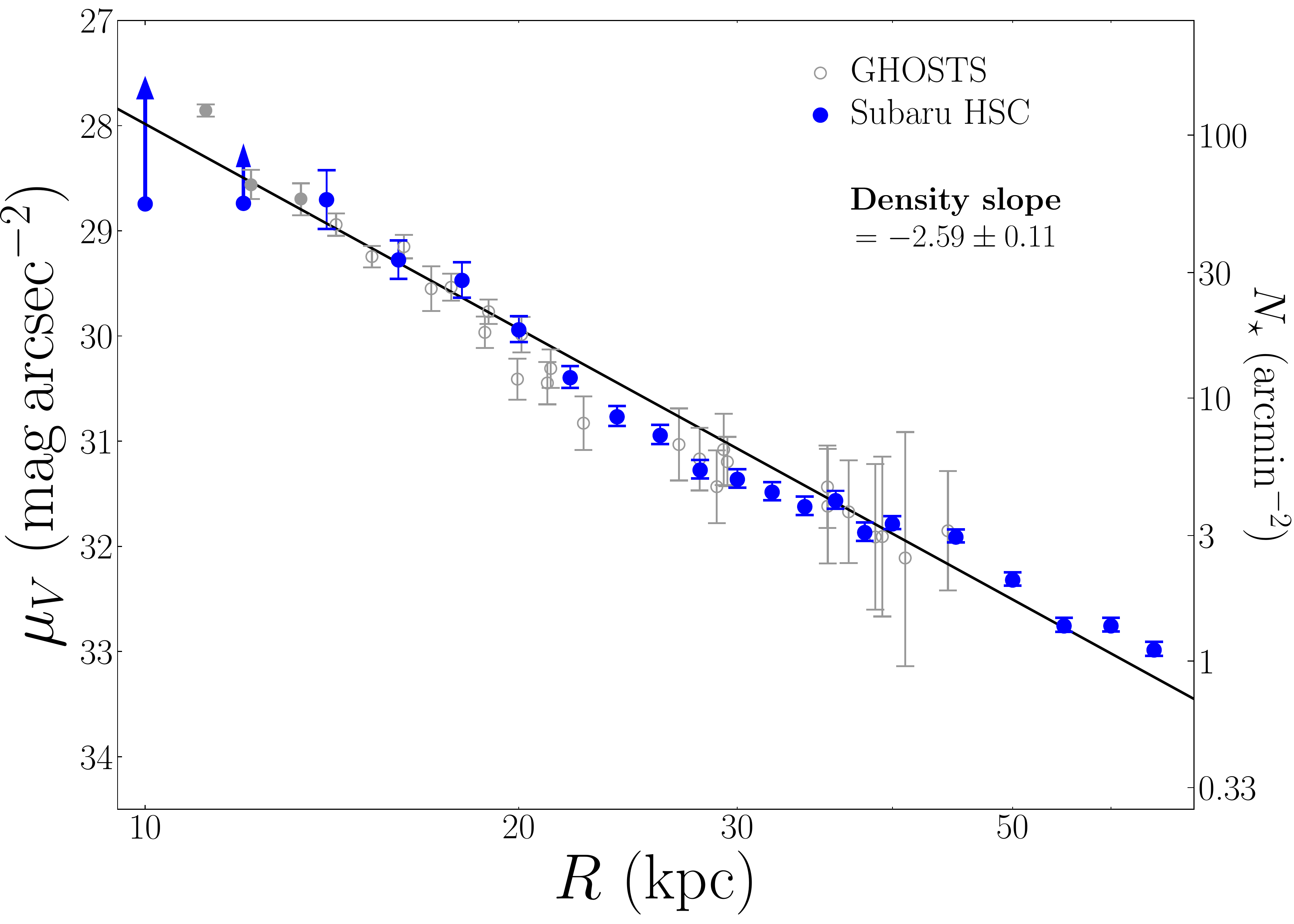}
\caption{M81's average minor axis SB profile (where SB is reported in $V$-band and radii in kpc) calculated from resolved star counts as described in \S\,\ref{sec:sb-prof}. The measurements made through this work are shown in blue, while measurements from the GHOSTS survey \citep{harmsen2017} are shown in gray for comparison. The three GHOSTS measurements used in the profile fit are shown as filled gray circles. All GHOSTS measurements have been corrected to exclude the initial background estimate as the minor axis CMDs show this to be an over-subtraction (see \S\,\ref{sec:sb-prof}). Corresponding star counts (stars per arcmin$^2$) are given on the right-hand y-axis. The solid black line is the best-fit density power-law to the data. The best-fit density slope is reported in the top right, which also agrees well with both the GHOSTS measurements. Reaching $\mu\,{\simeq}\,33$\,\magsqarc\ at 65\,kpc, this profile is \textit{one of the deepest ever measured}.}
\label{fig:minax-sb}
\end{figure*}

We caution that without similar extensive overlap with high-quality \textit{HST}-derived stellar catalogs, it is impossible to estimate the contribution from higher-metallicity stellar populations and, thus, this `blue-bias' is \textit{unable to be reliably corrected for}. This effect will present an issue for all similarly-designed ground-based stellar population surveys at distances $\gg$1\,Mpc. 

\section{Results}
\label{sec:results}
In this section, we first present quantitative measurements along M81's minor axis, including average surface brightness (SB) and $g{-}i$\ color profiles (given in Table \ref{tab:minax} of the \textsc{appendix}). We then present our results for the global stellar halo, including a map of resolved RGB stars, as well as a census of stellar mass in the M81 Group, including the contribution of tidal debris to the stellar halo. 

\subsection{The Minor Axis: Estimating M81's Past Accretion History}
\label{sec:minax}
The minor axes of galaxy halos are predicted to be relatively free of contamination by \textit{in situ} stars (generally defined as stars which were formed in the central galactic potential, rather than accreted; e.g., \citealt{pillepich2015} and references therein) beyond 10\,kpc \citep{monachesi2016b}. As M81 is a highly-inclined galaxy (inclination = 62\textdegree; \citealt{karachentsev2013}), its projected minor axis should be relatively free of such \textit{in situ} stellar populations, allowing minor axis measurements to directly trace the accreted stellar populations. As its current interaction appears to still be in its early stages, M81's minor axis is also relatively free of `contamination' from the debris of M82 and NGC 3077 \citep[e.g.,][Fig.\,\ref{fig:field-overview}]{okamoto2015}. We discuss the properties and impact of accounting for this debris in \S\,\ref{sec:global}. Thus, M81 is in a unique stage, where despite its ongoing interaction, its minor axis provides a reliable window onto its past ($\gtrsim$1\,Gyr ago) accretion history. Figure \ref{fig:minax-cmds} shows the stellar-selected (e.g., \S\,\ref{sec:stargal}) CMDs along M81's minor axis in nine broad radial bins out to 75\,kpc. Figure \ref{fig:minax-sb} shows the measured average SB and $g{-}i$\ color profiles along M81's minor axis. Their derivations are described in \S\,\ref{sec:sb-prof} and \S\,\ref{sec:color-prof}, respectively.

\subsubsection{Surface Brightness Profile}
\label{sec:sb-prof}
We define the minor axis according to the region shown in Figure \ref{fig:field-overview} in red. Leveraging our large survey footprint, we define a much wider minor axis region than is covered by the GHOSTS survey, allowing for more robust averaging and inclusion of any potential substructure absent in the sparse GHOSTS measurements. 

We divide the minor axis into projected annular radial regions, 2\,kpc wide from 10--40\,kpc, and wider 5\,kpc bins outside 40\,kpc, to account for the lower number of sources. We then calculate the average density for each radial region. Visually inspecting the CMDs in each bin, we find that at radii $>$65\,kpc along the minor axis, the RGB was indistinguishable from a CMD composed of only background galaxies and MW foreground stars (see Figure \ref{fig:minax-cmds}). We thus consider the halo beyond 65\,kpc along the minor axis to be undetected, and parameterize the density of RGB candidates at $>$65\,kpc as a uniform background equivalent to 0.53\,arcmin$^{-2}$. This is different from the findings of \cite{harmsen2017}, where it was assumed that stars outside of 45\,kpc were background contamination, due to the limited radial range and overall number counts. Using the extended radial range provided by HSC, as well as the improved star counts provided by our broad minor axis selection region, we detect RGB halo sources out to 65\,kpc, indicating that the initial GHOSTS SB profile was over-subtracted.

In order to correct our measured RGB counts for this background contamination, as well as intrinsic incompleteness due to photometric detection and source crowding, we turn to the results of our ASTs, described in \S\,\ref{sec:ast}. Using our discovery of the non-linearity between intrinsic and recovered RGB counts and the presence of a uniform contaminating background beyond 65\,kpc, we model `corrected' RGB density as a function of observed density, with the form
\begin{equation}
\label{eq:2}
    \Sigma_{\rm corr} = \Sigma^{*^\prime}\,(\Sigma_{\rm obs}\,{-}\,\Sigma_{\rm BG})\,e^{\beta\,(\frac{\Sigma_{\rm obs}\,{-}\,\Sigma_{\rm BG}}{\Sigma_0})^{\gamma}}.
\end{equation}
The best fit to our AST results, shown in Figure \ref{fig:ast-ghosts-density}, gives parameters of $\Sigma^{*^\prime}\,{=}\,2.09$, $\Sigma_{\rm BG}\,{=}\,0.53$, $\Sigma_{0}\,{=}\,140$, $\beta\,{=}\,2$, and $\gamma\,{=}\,1.6$. To corroborate the validity of this model, we also compare the RGB densities in each of the GHOSTS fields, measured with Subaru, to the densities published in \cite{harmsen2017}. Shown in Figure \ref{fig:ast-ghosts-density}, we find that after correcting for the over-subtraction of background sources, the GHOSTS measurements agree excellently with our model for corrected RGB counts. This is powerful and completely empirical support for our chosen model and AST approach. 

We calculate the mean RGB-selected source density in each radial region and then correct to the `intrinsic' density, accounting for background contamination and incompleteness, using Equation \ref{eq:2}. We then convert this corrected density to equivalent $V$-band surface brightness following \cite{harmsen2017}, as:
\begin{equation}
    \mu_{V} = {-}2.5 \log_{10}[\Sigma_{\rm corr}({\rm arcsec}^{-2}) \cdot 2.09{\times}10^{-10}], 
\label{eq:3}
\end{equation}
assuming a 10\,Gyr, [M/H]\,${=}\,{-}$1.2 isochrone model.

\begin{figure}[!t]
\leavevmode
\centering
\includegraphics[width={\linewidth}]{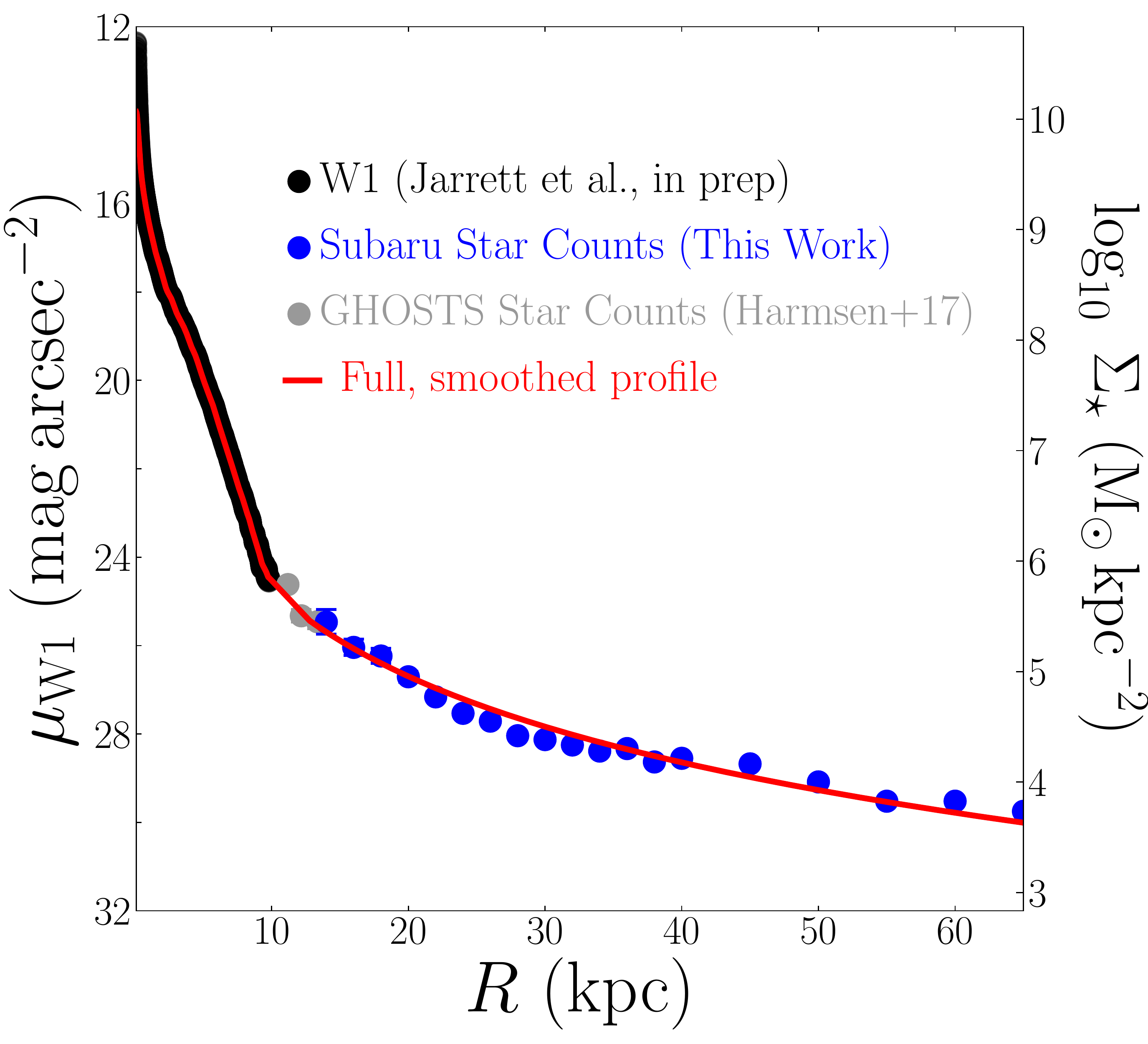}
\caption{Near-infrared SB profile along M81's minor axis, combining \textit{WISE} $W1$\ \citep{jarrett2019}, which probes M81's interior, with the outer resolved star profile obtained from this work. Corresponding stellar mass density is shown on the right axis (see \S\,\ref{sec:global} for conversion of $\mu_{W1}$\ to $\Sigma_{\star}$). Star counts have been converted to \textit{W1} using our adopted fiducial isochrone model (10\,Gyr, [Fe/H]\,${=}\,{-}1.2$; see \S\,\ref{sec:sb-prof}). Black points show the \textit{W1} measurements, while blue points show this work and the three gray points show inner GHOSTS measurements. A smooth, integrated profile is fit to the total profile and shown in red, for visual effect.}
\label{fig:full-minax}
\end{figure}

Uncertainties on the density measurements were carefully accounted for from three distinct sources. we assume errors on the average density in each radial section by taking the standard deviation in density across all independent $\sim$arcmin$^2$\ regions, divided by the square root of the number of these regions. Owing to the wide minor axis selection area, and therefore large number of independent arcmin$^2$\ regions in each radial section, these uncertainties are quite small --- ranging from $\sim$20\% at the smallest radii, to $\sim$1\% at the largest radii. Second, we account for the model-based correction of RGB density from our ASTs. Owing to the smoothness of the AST results, this is a small effect. As the high-density exponential portion of the model is both the most difficult to fit and has the greatest potential to impact mass estimates, we estimate the uncertainty by performing the fit while excluding either the highest-density or second highest-density point. The difference, which we take into account, is small --- approximately 1--5\%. Lastly, we calculate Poisson uncertainties on the total number of stars in each radial bin. Owing to the angular broadening of our minor axis selection region, as well as the broadening of the radial binning, with radius, these uncertainties are relatively constant at 3--5\%. 

Figure \ref{fig:minax-sb} shows the minor axis SB profile for M81. Our measurements are shown in blue, with the GHOSTS points shown in gray for comparison. Our measurements extend $\sim$50\% farther along M81's minor axis than GHOSTS and reach remarkable depths of $\mu_{V}\,{\simeq}$\,33\,\magsqarc\ at 65\,kpc. This is among the deepest SB profiles ever measured (e.g., compare to: $\mu_{V}\,{\sim}$\,32\,\magsqarc, PISCeS Survey, \citealt{crnojevic2016}; $\mu_{V}\,{\sim}$\,30\,\magsqarc, Dragonfly Survey, \citealt{merritt2016}).

To quantify the halo mass, we fit a power-law to the density profile. As Figure \ref{fig:minax-cmds} and \ref{fig:minax-sb} show, the inner two radial bins are impacted by crowding too severe to be accounted for by our AST-based corrective model. To account for this in our density fit, we combine our Subaru measurements from 15--65\,kpc with the three lower-radius GHOSTS measurements from \cite{harmsen2017} (11.2, 12.2, and 13.4\,kpc; shown as filled gray circles in Figure \ref{fig:minax-sb}) to fit the full density profile. We use a model of the form 
\begin{equation}
    \log_{10} \Sigma(R) = \log_{10} \Sigma_0 - \alpha \log_{10}(R/R_0),
\end{equation}
where the densities are expressed in arcsec$^{-2}$. This best-fit model yields parameters of $\log_{10}\Sigma_0$\ = $-$0.53, $R_0$\ = 4.19\,kpc, and $\alpha$\ = $-$2.59. Though this is much shallower than the density profile found by \cite{harmsen2017} ($\alpha$\ = $-$3.53), it is in good agreement with the GHOSTS measurements when corrected for their initially over-subtracted background (above). Following \cite{harmsen2017}, we integrate the profile from 10--40\,kpc, using elliptical annuli with the same assumed projected axis ratio of 0.61, obtaining an accreted stellar mass from 10--40\,kpc of $M_{\star,10{-}40}\,{=}\,3.7{\times}10^8\,M_{\odot}$. Extrapolating to total accreted mass using the \cite{harmsen2017} 10--40-to-total ratio of 0.32, we estimate a total accreted mass of $M_{\rm \star,Acc}\,{=}\,1.16{\times}10^9\,M_{\odot}$\ --- within 2\% of the GHOSTS estimate of $1.14{\times}10^9\,M_{\odot}$. Despite the change in power law parameters, the large characteristic radius, combined with the relatively narrow 10--40\,kpc radial range over which the profile is evaluated, results in nearly identical total mass estimates.

Finally, we compare our resolved star-based minor axis SB profile to integrated light measurements, which excel in the bright innermost parts of the galaxy, where resolved star measurements suffer from strong crowding. Figure \ref{fig:full-minax} combines our measured profile with a near-infrared version of M81's minor axis SB profile, following \cite{harmsen2017}. In this case, we have chosen the \textit{WISE W1} (3.4\,\textmu m) profile measured as part of the \textit{WISE} Enhanced Resolution Galaxy Atlas (\citealt{jarrett2012}; \citealt{jarrett2013}; T.H. Jarrett, private communication; \citealt{jarrett2019}). We have adjusted the elliptically-averaged profile to a minor axis-only version using the measured axis ratio for each elliptical annulus. Then, using the same 10\,Gyr, [Fe/H]\,${=}\,{-}1.2$\ isochrone model which was used to convert our RGB counts to $\mu_{V}$, we instead convert these counts to $W1$. The \textit{WISE} profile agrees well with our resolved star-based profile, with the the different methods converging nicely at 10\,kpc. 

\begin{figure*}[!ht]
\leavevmode
\centering
\includegraphics[width={0.8\linewidth}]{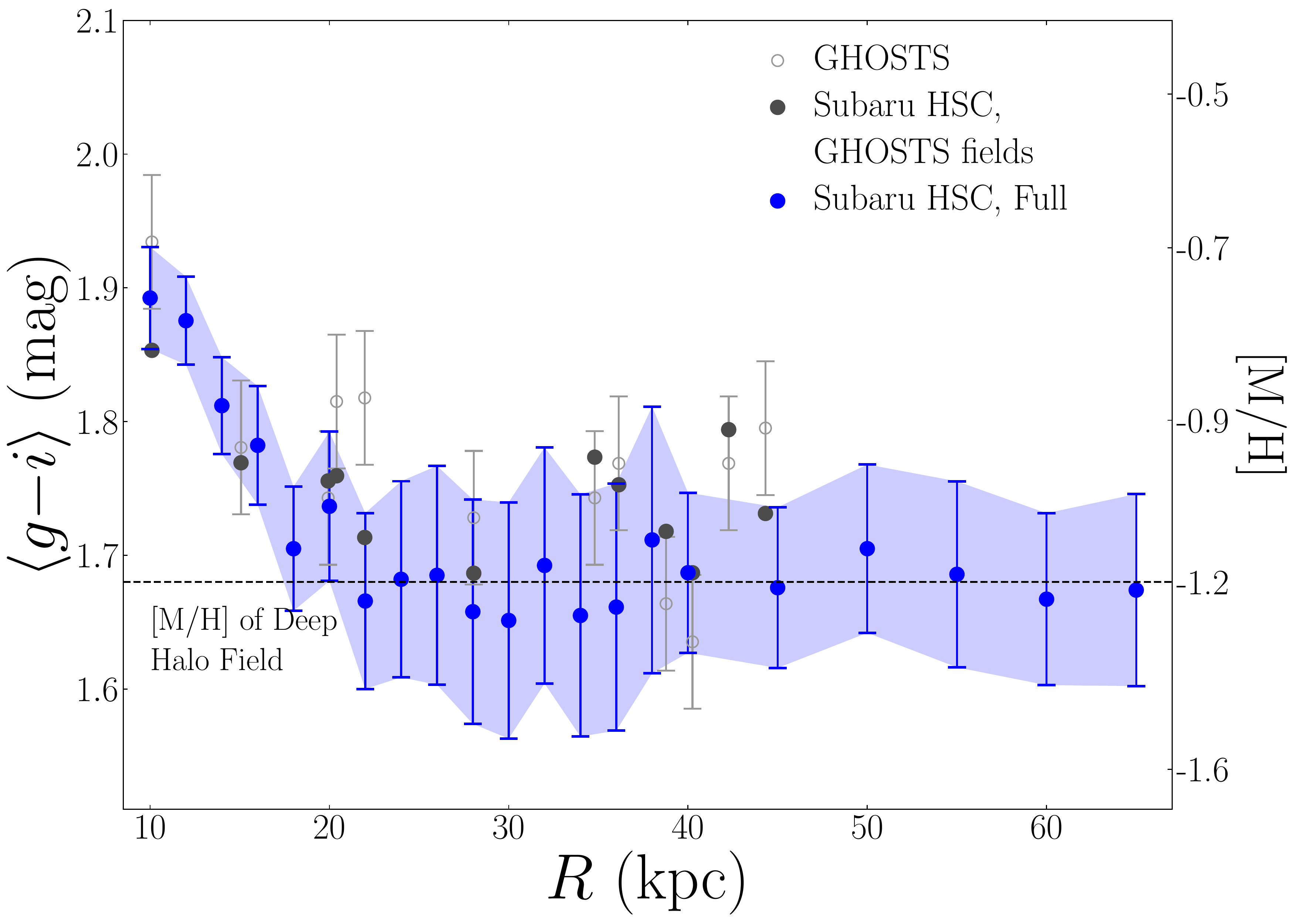}
\caption{Average $g{-}i$\ color profile of resolved RGB stars along M81's minor axis, as described in \S\,\ref{sec:color-prof}. Subaru HSC measurements are again shown in blue, while GHOSTS measurements \citep{monachesi2016a} are shown in light gray. Subaru measurements within the GHOSTS field areas, rather than our full minor axis region, are shown as dark gray points and display excellent agreement with the GHOSTS profile. Metallicity, calculated from equivalent F606W$-$F814 color \citep{streich2014}, is shown along the righthand $y$-axis. Additionally, we show the [M/H]\,=\,$-$1.2 metallicity measurement (dashed line) of M81's halo estimated from deep \textit{HST} data (reaching the Red Clump; \citealt{durrell2010}). We observe an extremely flat outer profile from 20--65\,kpc. We also resolve, for the first time, a distinct break in the color profile at $R\,{\lesssim}\,20$\,kpc, inside which the profile rises steeply --- $\sim$0.22\,mag in color, $\sim$\,0.51\,dex in metallicity from 10--20\,kpc.}
\label{fig:minax-color}
\end{figure*}

\subsubsection{Color Profile}
\label{sec:color-prof}
We calculate the average minor axis $g{-}i$\ color profile using the same minor axis area and radial regions as used for the SB profile (\S\,\ref{sec:sb-prof}). For detected sources in each radial bin, we convert the measured $g{-}i$\ color to $Q_{\rm Col}$\ by rotating the CMD $-$22\textdegree\ around a point (1.62,24.8) 0.5\,mag below the TRGB (see \S\,\ref{sec:ghosts-cal}; Figure \ref{fig:ghosts-qcol}). We also compute $Q_{\rm Col}$\ for all sources at $>$65\,kpc, assuming these to be background contamination (following Figure \ref{fig:minax-cmds}, \S\,\ref{sec:sb-prof}). For each radial bin, we scale these uniform background counts to the area of the bin. We then subtract the scaled cumulative $Q_{\rm Col}$\ distribution of these background sources from the measured cumulative $Q_{\rm Col}$\ distribution in each radial bin. The average $Q_{\rm Col}$\ at each radius is then calculated as the median of the background-subtracted distribution, which is then rotated back to obtain the average $g{-}i$. Following the results of our ASTs (see \S\,\ref{sec:ghosts-cal}), we add a constant 0.08\,mag to the average color at each radius, to account for the HSC observations' lacking sensitivity to red stellar populations observed in all GHOSTS fields.

We accounted for uncertainties on our average color measurements from two sources. We first estimate the upper and lower standard error around the average $Q_{\rm Col}$, by separately calculating the 16--50\% and 50--84\% percentile spreads of the $Q_{\rm Col}$\ distribution in each radial bin, and divide by the square root of the number of stars in each bin. Second, we estimate standard $\sqrt{N}/N$\ Poisson errors at each radius, where $N$\ is the number of stars in each bin.

Figure \ref{fig:minax-color} shows the minor axis color profile for M81. Our measurements are again shown in blue, and the GHOSTS points again in light gray for comparison \citep{monachesi2016a}. While in relative agreement with the GHOSTS measurements, our Subaru profile appears to be $\sim$0.05\,mag bluer than the GHOSTS profile, though with much less scatter. To ascertain the origin of this discrepancy, we measured the average color in each of the overlapping GHOSTS field regions. Though there is considerable scatter, due to the low number of RGB-like sources detected in regions as small as the ACS and WFC3 fields, we confirm that, on average, the median colors measured with Subaru in each GHOSTS field exhibit the same `blue bias' observed in the ASTs (\S\,\ref{sec:ast}). The average Subaru-measured colors in each GHOSTS field, corrected for the 0.08\,mag `blue bias' between Subaru and GHOSTS, are shown in dark gray. These measurements agree much better with the original GHOSTS measurements, suggesting that the difference in the profiles is due to GHOSTS narrow and more stochastic coverage of M81's minor axis, compared to the global view provided by Subaru.

We recover the GHOSTS measurement of a $\sim$flat profile at $R\,{\gtrsim}\,20$\,kpc, $g{-}i\,{\simeq}$\,1.68. However, we also observe a distinct negative color gradient for $R\,{\lesssim}\,20$\,kpc, which is not due to the effects of crowding, which we observe to have no color-dependence in our AST analysis (\S\,\ref{sec:ast}). This gradient smoothly connects the flat region of the profile to a single inner GHOSTS field (10\,kpc), observed by \cite{monachesi2013,monachesi2016a}, which is much redder than the outer fields. At first seemingly a `jump' in the profile, when combined with our Subaru observations, this inner field measurement appears to confirm that M81 possesses a steep minor axis color gradient within 20\,kpc.

To estimate how this translates to metallicity, we use model \textit{HST}--SDSS color--color tracks (\S\,\ref{sec:ghosts-cal}) to convert our average $g{-}i$\ colors to metallicity, using the calibration of \cite{streich2014}. Though this conversion is somewhat uncertain, it is heartening that the outer portion (i.e. $>$20\,kpc) of our halo profile matches the previous estimate of [M/H]\,=\,$-$1.2 \citep{durrell2010,monachesi2013}, which used deep \textit{HST} data reaching the `Red Clump', almost exactly. With this metallicity calibration, we estimate that the $\sim$0.22\,mag change in color from 10--20\,kpc corresponds to a $\sim$0.51\,dex change in [M/H], from ${\sim}\,{-}$1.2 to ${\sim}\,{-}$0.69. This yields a metallicity gradient of slope ${\sim}\,{-}$0.05\,dex\,kpc$^{-1}$\ inside 25\,kpc --- 5$\times$\ steeper than the global metallicity profile of M31, and comparable to M31's inner 25\,kpc \citep{gilbert2014}.

While this is the first observed case of such a distinct break in the color/metallicity profile of a MW-mass galaxy, galaxies with similar metallicity profiles to M81 --- i.e. displaying negative initial gradients, which flatten at large radii --- have been observed in simulations \citep[e.g.,][]{monachesi2019}. However, it is very rare to find even a simulated galaxy with such a sharp transition at $<$\,30\,kpc. We discuss two possible origins of this steep color profile in \S\,\ref{sec:saga}.

\begin{figure*}[!ht]
\leavevmode
\centering
\includegraphics[width={0.85\linewidth}]{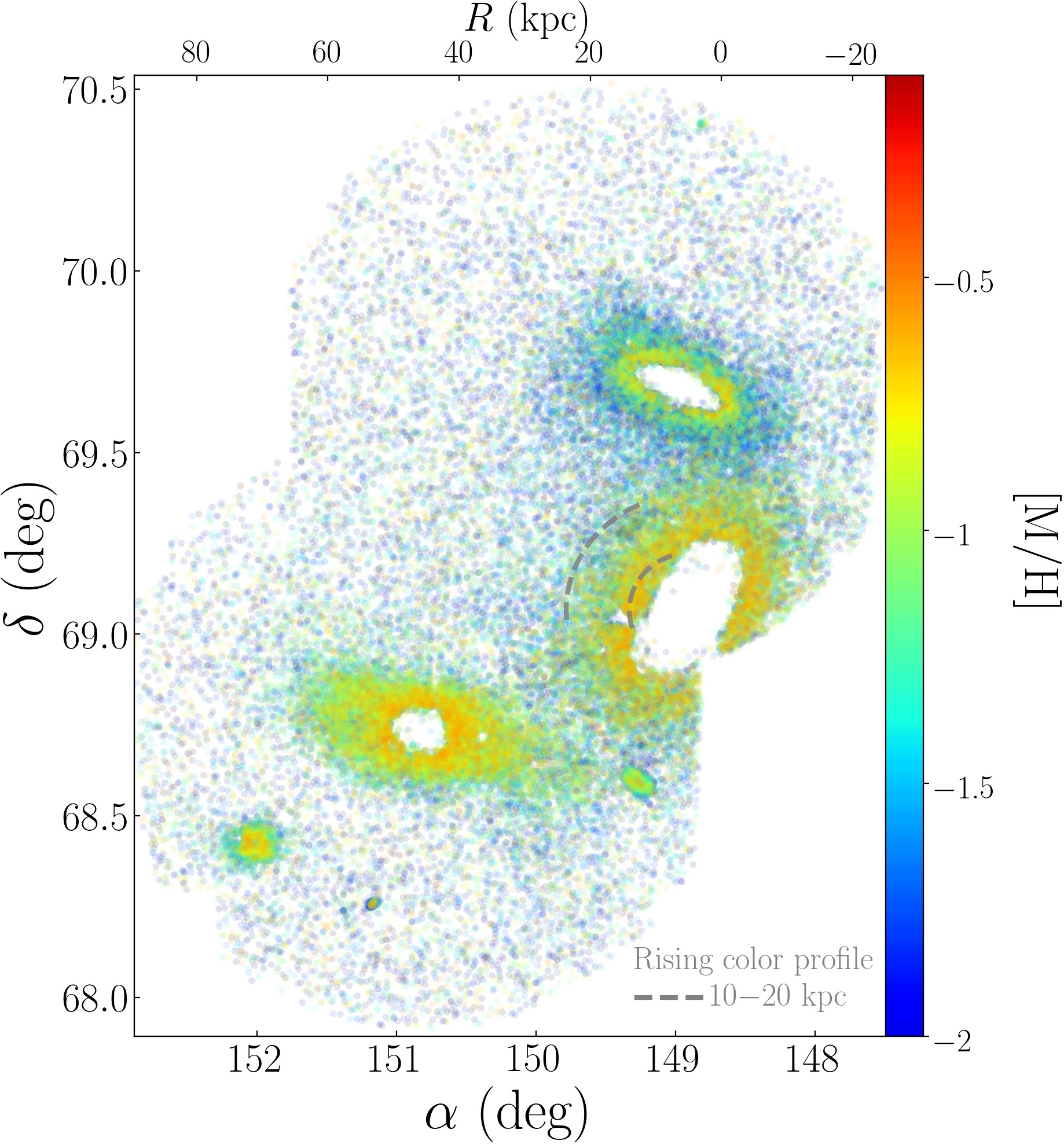}
\caption{Map of resolved RGB stars in the stellar halo of M81. Points have been color-coded by metallicity, determined from isochrone fitting (\S\,\ref{sec:global}). A scale bar giving projected distance from M81 is shown along the top $x$-axis. Two annular sections, showing the 10--20\,kpc radial range within which we measure a rising color profile in Figure \ref{fig:minax-color}, are shown as gray dashed curves. The metal-rich debris from the triple-interaction visually dominates against the surrounding metal-poor halo, though the minor axis remains clear of this debris.}  
\label{fig:rgb}
\end{figure*}

\subsection{The Global Stellar Halo of M81}
\label{sec:global} 
While M81's minor axis is a window onto its past accretion history, the global halo properties provide a window onto the current interaction. We first present the globally resolved populations in M81's halo and conduct a census of stellar mass (\S\,\ref{sec:resolved-mass}), followed by an accounting of the tidal debris around M82 and NGC 3077, and how it impacts M81's current halo properties (\S\,\ref{sec:tidal}). 

\begin{figure*}[!ht]
\leavevmode
\centering
\includegraphics[width={0.85\linewidth}]{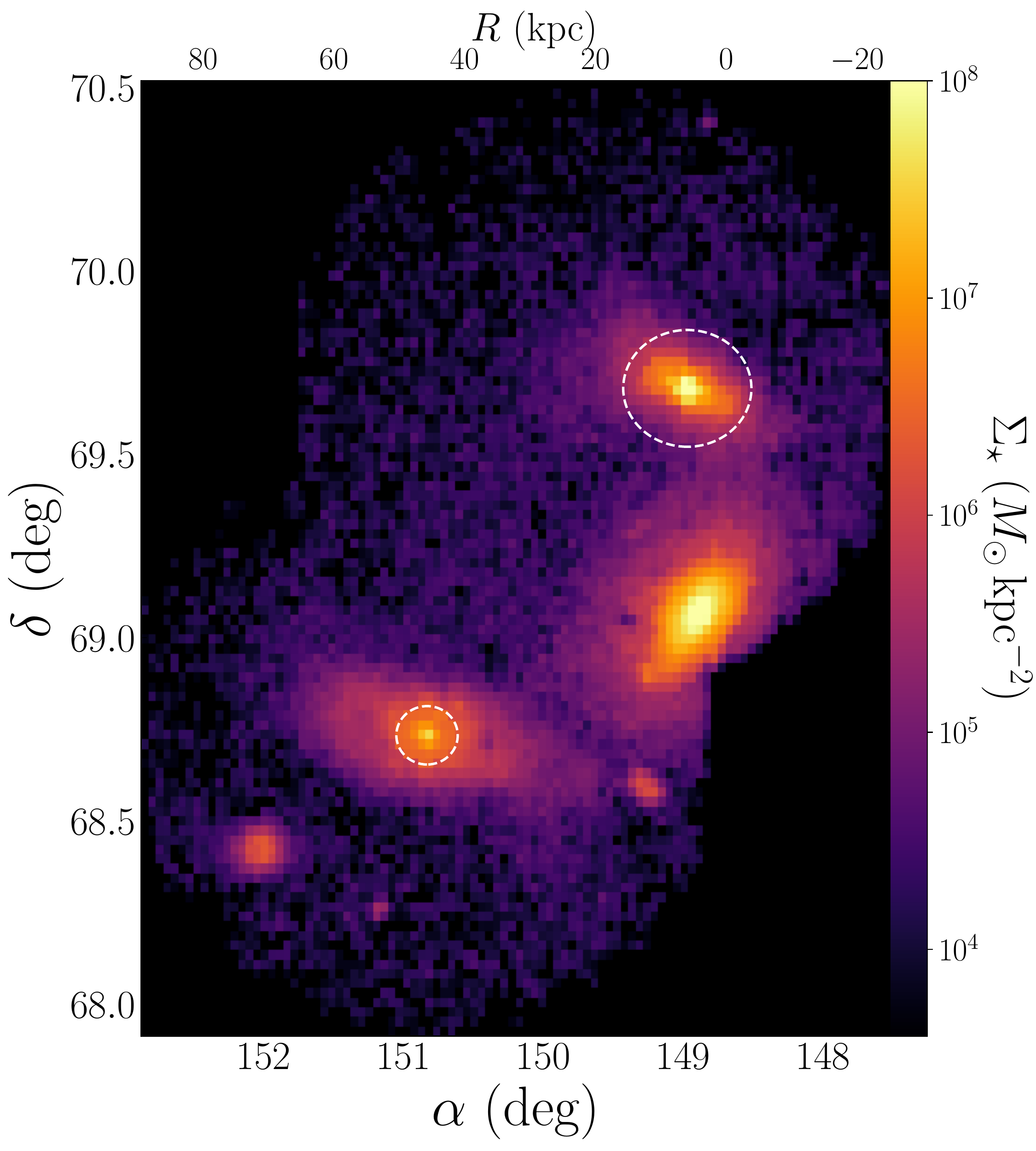}
\caption{Stellar mass density map of the M81 Group. The map has been logarithmically scaledand color-coded according to the bar on the right. Density was calculated for each $\sim$1\,kpc$^{2}$\ pixel, and converted to stellar mass according to \S\,\ref{sec:denscal} and \S\,\ref{sec:resolved-mass}. The interior regions of M81, M82, and NGC 3077, where the data were too crowded to detect individual stars with Subaru (see Figure\,\ref{fig:rgb}), were filled in using calibrated $K_{\rm s}$\ images from the 2MASS Large Galaxy Atlas \citep{jarrett2003}, which were re-binned to $\sim$1\,kpc physical resolution. The final map was lightly smoothed with a 0.5\,kpc Gaussian kernel. The final map spans an impressive four\,orders of magnitude in mass density. White dashed circles show the estimated tidal radii of M82 and NGC 3077. We count all material outside of these circles as unbound, to estimate the total current accreted mass of M81 (\S\,\ref{sec:tidal})}.
\label{fig:density}
\end{figure*}

\subsubsection{Stellar Populations and Stellar Mass}
\label{sec:resolved-mass}
In Figure \ref{fig:rgb} we present a global map of resolved RGB stars in M81's halo. Each star has been color-coded by its best-fit photometric metallicity, rather than $g{-}i$\ color, as metallicity is the more intuitive (while uncertain) quantity, and is more directly comparable to other similar datasets. For this result, we estimate metallicity for each individual star, using a grid of PARSEC isochrones \citep{bressan2012}, Age\,=\,10\,Gyr, ranging from [M/H]\,=\,$-$2 to 0 with steps of $\Delta$[M/H]\,=\,0.05\,dex. The distance in $g{-}i$\ color, at the given $i$\ magnitude, is evaluated for each star, for each isochrone. The best-fit metallicity is then defined as the model which minimizes the data$-$model $g{-}i$\ color residual. We display [M/H], rather than [Fe/H], so as to remain agnostic about [$\alpha$/Fe]. Accounting for photometric uncertainties alone (not systematic uncertainties associated with age or different stellar evolution models), the typical [Fe/H] error is $\leqslant$\,0.2\,dex.

The ongoing interaction between M81, M82, and NGC 3077 is immediately visible in the resolved star map. NGC 3077 outskirts display an `S' shape, typical in tidally disrupting systems, while M82's debris is more compact. The tidal debris around both satellites is quite metal-rich. The rest of the halo, however, is quite metal-poor, comparable to M81's minor axis. Other than the interaction debris, five previously-known satellite galaxies are visible (see Figure \ref{fig:field-overview}; IKN: \citealt{karachentsev2006}; BK5N: \citealt{caldwell1998}; KDG 61: \citealt{karachentseva&karachentsev1998}; d0955+70: \citealt{chiboucas2009,chiboucas2013}; d1005+68: \citealt{smercina2017}), though there are no obvious substructures. 

Figure \ref{fig:density} turns our map of resolved RGB stars into a map of stellar mass density in M81's halo. Using the method described in \S\,\ref{sec:denscal}, we convert our RGB map to corrected RGB counts (Equation \ref{eq:2}). Again using a fiducial Age\,=\,10\,Gyr, [Fe/H]\,$-$1.2 isochrone (following \citealt{harmsen2017}), we convert RGB density to a  corresponding stellar mass density, $\Sigma_{\star}$\ in $M_{\odot}\,{\rm kpc}^{-2}$, computed within $\sim$kpc\,$\times$\,kpc pixels. We showed in Figure \ref{fig:full-minax} that this method of SB/stellar mass estimation agrees well with integrated light measurements. The crowded centers of M81, M82, and NGC 3077 (see Figure \ref{fig:rgb}) have been filled in with publicly available $K_{\rm s}$-band images from the 2MASS Large Galaxy Atlas \citep{jarrett2003}. We have clipped $\Sigma_{\star}$\ to $>$5.4$\times$10$^3\,M_{\odot}\,{\rm kpc}^{-2}$\ --- equivalent to $\mu_V\,{\simeq}$\,33\,\magsqarc, or one RGB star\,kpc$^{-2}$, the faintest limit we measure along the minor axis (e.g., Figure \ref{fig:minax-sb}, \S\,\ref{sec:sb-prof}). Combining our star count measurements with traditional near-infrared imaging, this map of stellar mass spans $>$4 orders of magnitude --- from the dense stellar bulges at the centers of the primary galaxies, to the faintest stellar outskirts. This is among the most sensitive maps of stellar mass-density ever constructed for a MW-mass galaxy. 

\begin{figure*}[!ht]
\leavevmode
\centering
\includegraphics[width={0.7\linewidth}]{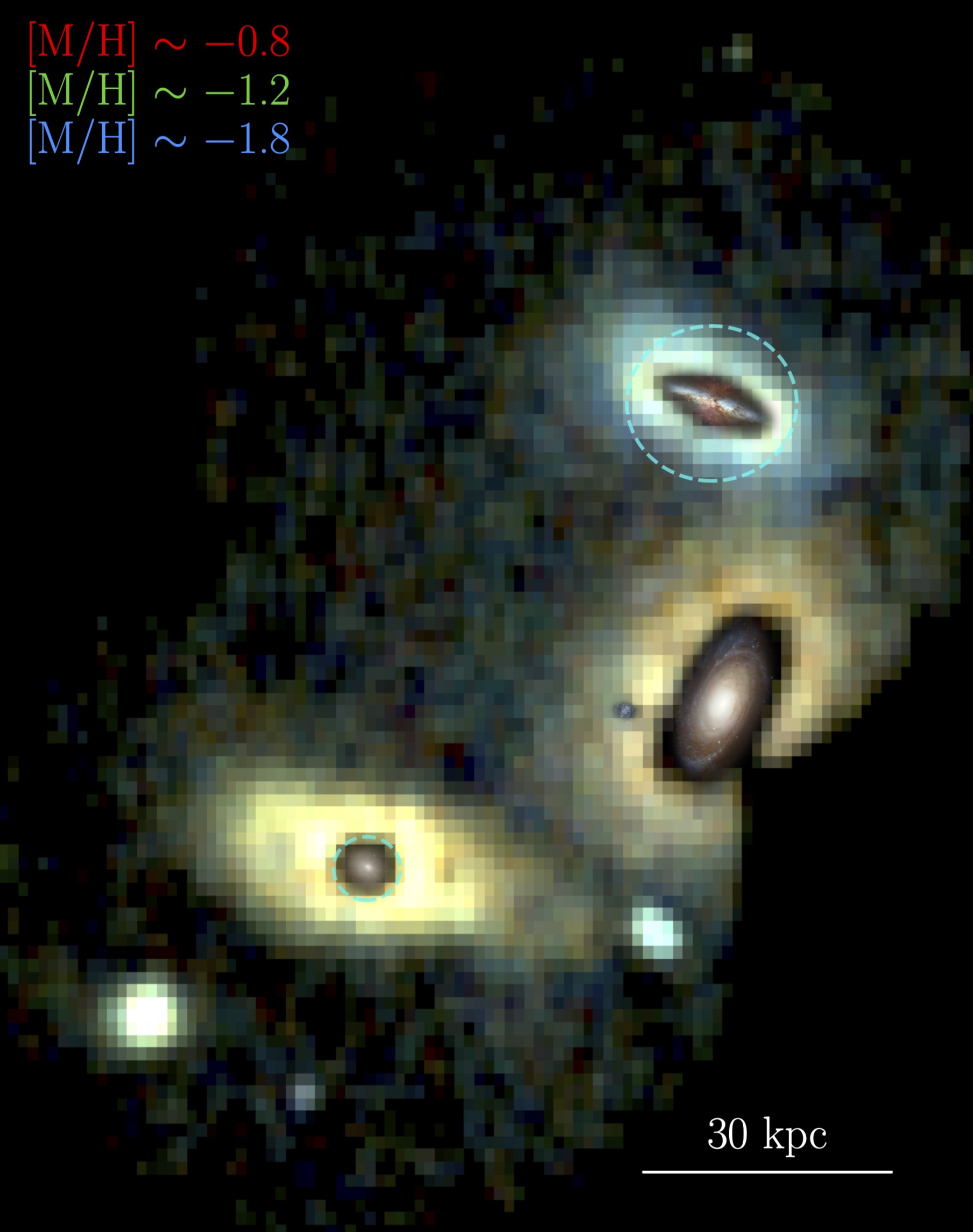}
\caption{Density image of RGB stars, with intensity mapped to stellar density, where each `channel' represents stars in three bins of metallicity: [Fe/H]\,${\sim}\,{-}$0.8 (red), [Fe/H]\,${\sim}\,{-}$1.2 (green), and [Fe/H]\,${\sim}\,{-}$1.8 (blue). Each channel was smoothed using first a tophat filter of size $\sim$20\,kpc (to bring out substructure), and then a Gaussian filter of width $\sim$1\,kpc. The interiors of M81, M82, and NGC 3077 have been filled with to-scale images from \textit{HST} (\uline{credit}: NASA, ESA, and the Hubble Heritage Team).}
\label{fig:channel}
\end{figure*}

\subsubsection{Tidal Debris Around M82 and NGC 3077}
\label{sec:tidal}
Figure \ref{fig:rgb} \& \ref{fig:density} clearly indicate that there is a significant amount of metal-rich stellar material around M82 and NGC 3077. While M81's minor axis gives the properties of its past accretion history (i.e. $M_{\rm Acc}\,{=}1.16\,{\times}10^9\,M_{\odot}$; see \S\,\ref{sec:sb-prof}), any of the material around the two satellites which is unbound should be included in the \textit{current} halo properties. To estimate how much of the material is unbound from M82 and NGC 3077, we estimate their respective tidal radii, using the basic approximation \citep{vonHorner1957,king1962},
\begin{equation}
    r_{\rm tid} \simeq R\,\left(\frac{M_{\rm \star,sat}}{2\,M_{\rm enc}(R)}\right)^{1/3},
\end{equation}
where $r_{\rm tid}$\ is the tidal radius, $R$\ is the separation between the central and the satellite adjusted for projection (i.e. $R\,=\,\sqrt{3}\,R_{proj}$), $M_{\rm \star,sat}$\ is the stellar mass of the satellite, and $M_{\rm enc}(R)$\ is the total mass of the central enclosed within $R$. To estimate $M_{\rm enc}(R)$, we adopt the familiar approximation for a flat rotation curve, 
\begin{equation}
    M_{\rm enc}(R) = \frac{v_{\rm c}^2\,R}{G}, 
\end{equation}
where we have taken $v_{\rm c}\,{=}\,230\,{\rm km\,s^{-1}}$\ from M81's H\,I rotation curve at 10\,kpc \citep{deblok2008}.

The projected separations from M81 of M82 and NGC 3077 are 39\,kpc and 48\,kpc, respectively, and their stellar masses are 2.8$\times$10$^{10}\,M_{\odot}$\ and 2.3$\times$10$^{9}\,M_{\odot}$\ (S4G; \citealt{sheth2010}, \citealt{querejeta2015}). Taking $v_{\rm c}\,{=}\,230\,{\rm km\,s^{-1}}$, this yields projected tidal radii of 10\,kpc for M82 and 8.2\,kpc for NGC 3077. Circles with radii equal to these tidal radii are shown in white on Figure \ref{fig:density}. We then consider all material outside of these circles to be unbound. This amounts to 2.12$\times$10$^8\,M_{\odot}$\ around M82 and 3.36$\times$10$^8\,M_{\odot}$\ around NGC 3077 --- a total of $\sim$5.4$\times$10$^8\,M_{\odot}$, which is a substantial fraction of M81's integral past accreted mass ($\sim$10$^9\,M_{\odot}$). Taking a mass-weighted average metallicity of this material yields [Fe/H]\,$\simeq$\,$-$0.9 --- significantly more metal-rich than the rest of the halo. 

Figure \ref{fig:channel} combines Figures \ref{fig:rgb} and \ref{fig:density}. The mass-density map is divided into three average metallicity channels: [Fe/H]\,$\sim$\,$-$0.8 (red), [Fe/H]\,$\sim$\,$-$1.2 (green), and [Fe/H]\,$\sim$\,$-$1.8 (blue). Each channel is then intensity-weighted and combined into a three-channel color image. This figure highlights the visual impact that the massive and metal-rich debris around M82 and NGC 3077 has on the inferred mass and metallicity of M81's halo.

\section{The Saga of M81}
\label{sec:saga}

\subsection{A Quiet History}
As discussed in \S\,\ref{sec:sb-prof}, the total accreted stellar mass from M81's past accretions is $M_{\star,Acc}\,{=}\,1.16{\times}10^9\,M_{\odot}$, and is quite metal-poor ([Fe/H]\ $\sim$\ $-$1.2). If we take the limit that a single satellite dominates the halo properties, then the relationship between stellar halo mass and the mass of the most dominant satellite from \cite{dsouza&bell2018a} suggests M81's largest past merger was \textit{at most} $M_{\star}\,{\sim}\,5{\times}10^8\,M_{\odot}$\ --- the mass of the Small Magellanic Cloud (SMC; \citealt{mcconnachie2012}). Further, though we cannot reliably constrain the origin of M81's inner color profile, if it has an accretion origin, the steepness of the slope ($\sim$0.05\,dex\,kpc$^{-1}$) suggests that the event likely occurred early in M81's life \citep{dsouza&bell2018a}. It is interesting to note that the MW shows tentative evidence for a rising metallicity profile inside 30\,kpc as well \citep{conroy2019}, though the 3-D measurements, aided by precise distances, are very different from the 2-D projected measurements presented here. 

\begin{figure*}
\leavevmode
\centering
\includegraphics[width={0.99\linewidth}]{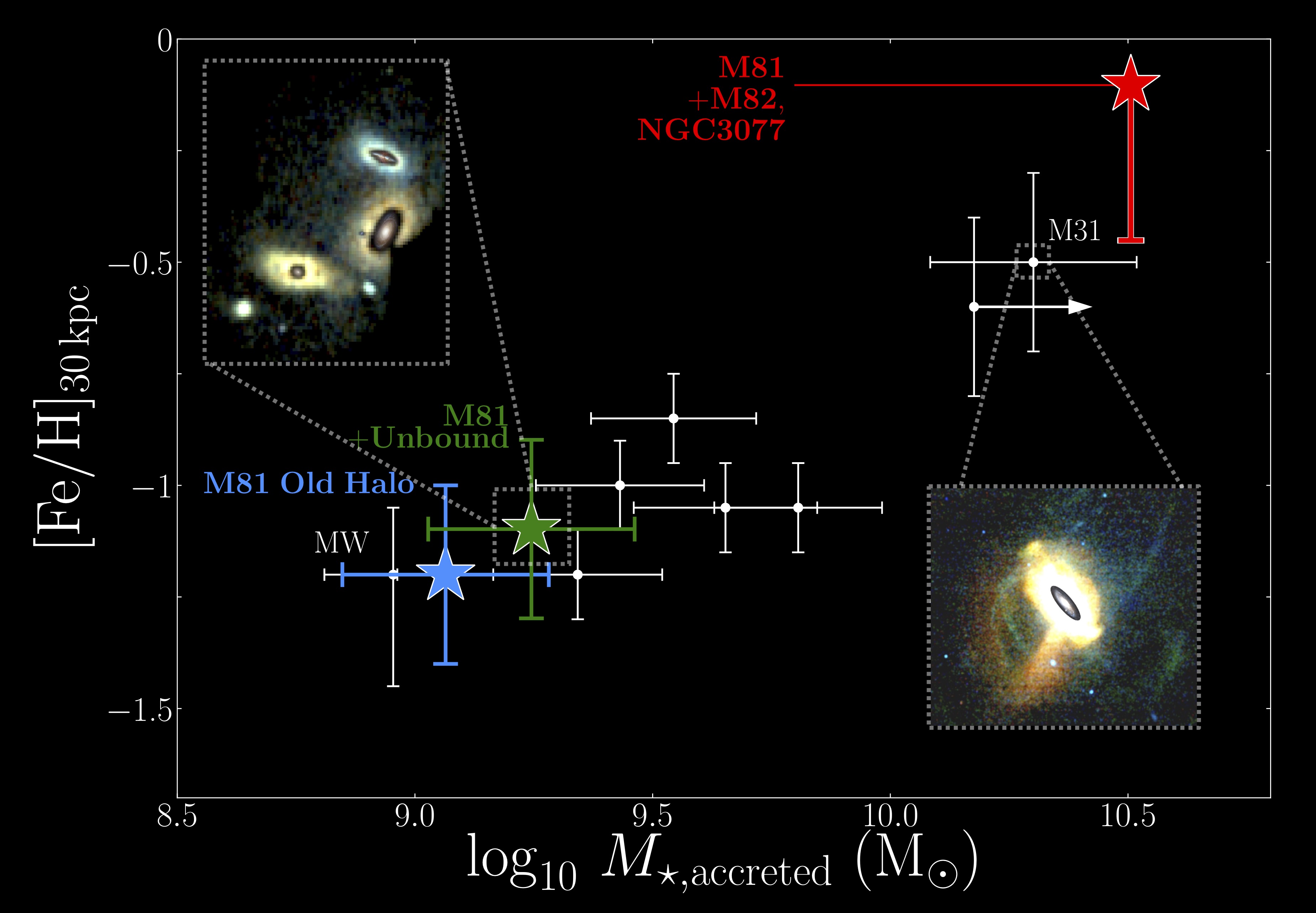}
\caption{The stellar halo mass--metallicity relation. Total accreted mass ($M_{\star,Acc}$) is plotted against metallicity measured at 30\,kpc ([Fe/H]$_{\rm 30\,kpc}$). The evolution of M81's stellar halo is shown at three points (large stars): (1) its past accretion history (\textcolor{blue!65!}{\bf blue}), measured from the minor axis (see \S\,\ref{sec:sb-prof} \& \ref{sec:color-prof}), (2) its `current' halo (\textcolor{green}{\bf green}), accounting for unbound tidal debris around M82 and NGC 3077 (see \S\,\ref{sec:tidal}), and (3) its estimated properties following the accretion of M82 and NGC 3077 (\textcolor{red}{\bf red}; see \S\,\ref{sec:halo-formation}). For comparison, nearby galaxies (taken from \citealt{bell2017}) are shown in white; the MW and M31 are labeled separately, to highlight their opposite positions on the relation. The MW's stellar halo mass and metallicity are taken from \cite{mackereth2019} and \cite{conroy2019}, respectively. We adopt 50\% larger error bars than intially reported for each, to reflect the substantial spread from other measurements \citep[e.g.,][]{bell2008,deason2019}. Metallicity-coded channel density maps are shown as zoomed insets for both M81 (e.g., see Figure \ref{fig:channel}) and M31 (PAndAS; \citealt{martin2013}) as visual guides of M81's potential halo evolution. For points (1) and (2) we adopt 50\% uncertainties on total accreted mass and 0.2\,dex uncertainties on metallicity, following \cite{harmsen2017}. For (3), the large error in metallicity indicates our uncertainty about the final metallicity gradient of the halo. In this case, the red star assumes the central metallicities for both M82 and NGC 3077 (mass-weighted), while the error bar shows the impact of assuming a steep halo metallicity gradient such as observed in M31 \citep{gilbert2014}. Dominated by the accreted material from M82, M81's halo will be transformed from low-mass and metal-poor, to a massive and metal-rich halo, rivaling that of M31.}
\label{fig:evolution}
\end{figure*}

If, instead, the color gradient is driven by increasing contribution of \textit{in situ} material at small radii \citep[e.g.,][]{zolotov2009,font2011,monachesi2016b,monachesi2019}, then M81's current accreted mass estimate is an upper limit. When the jump in M81's color profile was initially observed, \cite{monachesi2013} suggested that the most likely explanation was \textit{in situ} material from M81's disk. To estimate the range of possible \textit{in situ} fractions, we assume the color of accreted material to be the average color of the `flat' part of the color profile --- $g{-}i\,{\simeq}\,1.68$. The average color (using the $Q$\ method described in \S\,\ref{sec:ghosts-cal} \&\ \S\,\ref{sec:color-prof}) of RGB stars in the center of M81 --- using a central \textit{HST} pointing from the GHOSTS survey (Field 01, $\sim$3\,kpc) --- is $g{-}i\,{=}\,2.17$, which we adopt as an upper limit on the `fiducial color' of the \textit{in situ} populations. Using the accreted ($f_{Acc}$) and \textit{in situ} ($f_{\rm IS}\,{=}\,1{-}f_{Acc}$) fractions as weights to produce the observed average color profile, we calculate $f_{Acc}$\ as a function of radius, and then convolve it with the observed density profile to estimate the integral change to estimated stellar halo mass. In the case of an \textit{in situ} origin for the steep inner color profile, we find a lower limit on the accreted fraction of $f_{Acc}\,{=}\,0.80$\ --- corresponding to a lower limit on M81's total accreted mass of $M_{\star,Acc}\,{=}\,9.3{\times}10^{8}\,M_{\odot}$. 

The punch line: regardless of the origin of its intriguing steep inner color profile, M81 has likely experienced a quiet accretion history for the vast majority of its life, accreting only satellites the size of the SMC or smaller.

\subsection{The Formation of a Massive Stellar Halo}
\label{sec:halo-formation}
That quiet history is over, however. M81 ($6.3{\times}10^{10}\,M_{\odot}$; \citealt{querejeta2015}) is currently undergoing a $\sim$1:2 merger with its massive satellite M82 ($2.8{\times}10^{10}\,M_{\odot}$; \citealt{querejeta2015}) and the $\sim$LMC-mass NGC 3077 ($2.3{\times}10^{9}\,M_{\odot}$; \citealt{querejeta2015}). In \S\,\ref{sec:tidal}, we showed that there is a significant amount of metal-rich material currently unbound from M82 and NGC 3077 --- $\sim$5.4$\times$10$^{8}\,M_{\odot}$, [Fe/H]\,$\simeq$\,$-$0.9. Accounting for this unbound material increases M81's average halo metallicity and increases M81's accreted mass by $\sim$50\%. It is clear from their star formation histories that M82 and NGC 3077 began their interaction with M81 \textit{at the same time}. Moreover, the star formation history of the group, including bursts of star formation in the disk of M82 \citep[e.g.,][]{rodriguez-merino2011,lim2013}, the center of NGC 3077 \citep[e.g.,][]{notni2004}, the tidal H\,I field between the three galaxies \citep[e.g.,][]{deMello2008}, and `tidal' dwarf galaxies such as Holmberg IX \citep[e.g.,][]{sabbi2008}, all suggest that this merger began $<$\,1\,Gyr ago. In $<$\,1\,Gyr this merger has already had a substantial impact on the properties of M81's stellar halo. 

Though a robust dynamical model does not exist for the future of the M81 system, such models have been constructed for the MW's interaction with the LMC. \cite{cautun2019} estimate that the LMC will merge with the MW within $\sim$2.4\,Gyr. Though the orbital properties of M82 and NGC 3077 are unclear, M82 is significantly more massive than the LMC, and thus will likely merge with M81 within the next $\sim$2\,Gyr. What, then, will be the properties of M81's stellar halo $\sim$2\,Gyr in the future, following its accretion of M82 and NGC 3077? The addition to the accreted mass is simply the combined stellar mass of both satellites --- an addition of $\sim$3$\times$10$^{10}\,M_{\odot}$\ (93\% comes from M82), which is $>$20$\times$\ larger than the total current accreted mass. Clearly this merger event will dominate the stellar halo mass of M81. The metallicity will also be significantly impacted. Assuming M82 and NGC 3077 follow the galaxy stellar mass--metallicity relation, they possess metallicities of [Fe/H]\,$\sim$\,0 and [Fe/H]\,$\sim$\,$-$0.6, respectively \citep{gallazzi2005} --- much higher than the stellar halo's current metallicity of [Fe/H]\,$\simeq$\,$-$1.2. 

In Figure \ref{fig:evolution}, we show the evolution of M81's stellar halo properties in the context of the observed stellar halo mass--metallicity relation for eight nearby MW-mass galaxies \citep[e.g.,][]{bell2017}, discussed in \S\,\ref{sec:intro}. Though several versions of this relation exist in the literature, here we adopt, as metrics, total accreted stellar mass ($M_{\star,Acc}$; \textit{x}-axis) and metallicity measured at 30\,kpc ([Fe/H]$_{\rm 30\,kpc}$; \textit{y}-axis). 

Prior to its current interaction, M81 possessed one of the lowest-mass and metal-poorest stellar halos in the nearby universe; among the eight examples shown here, only the MW is comparable in mass and metallicity. The massive tidal debris from M82 and NGC 3077 augments and enriches its stellar halo, but \textit{rapidly}. This is no modest evolution of halo properties, but an initial step precipitating a giant leap. In the next several Gyrs, after the merger has completed, the enormous amount ($M_{\star}\,{\simeq}\,3{\times}10^{10}\,M_{\odot}$) of metal-rich material accreted from M82 and NGC 3077 ([Fe/H]\,$\sim$\,$-$0.1; mass-weighted material from both M82 and NGC 3077) will have completely transformed M81's stellar halo --- the resulting behemoth will have few peers in the nearby Universe. Among its few rivals will be well-known examples of massive stellar halos such as Cen A, NGC 3115, and the stellar halo paragon: M31. In fact, in stellar mass, central density, and starbursting nature, M82 strongly resembles the proposed progenitor galaxy M32p, which \cite{dsouza&bell2018b} hypothesize merged with M31 $\sim$2\,Gyr ago, resulting in M31's current massive stellar halo. Though M81's halo will resemble these others in its broad properties, we have little idea what the density and color gradients, or resultant substructure, of the emergent halo will look like. Comparisons of these more nuanced halo metrics, while important, is beyond the scope of this work.

\textit{This is the first complete view of the evolution of a galaxy's stellar halo throughout a merger event}. It is clear that such a window on a major merger event has the potential to help us better understand the formation and evolution of systems with massive stellar halos, such as M31. Between the measurements along M81's minor axis and the analysis of its current merger with M82 and NGC 3077, we have constrained M81's three largest merger partners over its lifetime: (1) M82, (2) NGC 3077 --- an LMC-analog, and (3) the ancient $\sim$SMC-mass primary progenitor of M81's past halo. If not for M82, M81's dominant merger history would closely resemble that of the MW. M81's ancient accreted halo is very comparable to the MW's halo (Figure \ref{fig:evolution}), indicating that a single stochastic, M82-like merger is capable of transforming a MW-like halo into a halo such as M31's. This is direct and powerful evidence that the diversity in stellar halo properties is thus driven primarily by the diversity in the properties of the most dominant mergers.

\section{Conclusions}
\label{sec:conclusions}
We have presented a survey of the stellar halo of M81 with Subaru HSC. Using a suite of ASTs, as well as abundant existing \textit{HST} fields from the GHOSTS survey, we have precisely calibrated and corrected our wide-field, ground-based catalog of RGB stars, in order to obtain one of the most detailed views of a stellar halo outside of the LG. We find the AST-based corrections, in concert with \textit{HST} data, to be \textit{crucial for measuring accurate stellar population properties}, and caution that without similar extensive overlap with space-based stellar catalogs, the color-dependent effects of completeness in any distant ($\gg$1\,Mpc), ground-based stellar halo measurements may be difficult to account for. We measure:

\begin{enumerate}
    \item M81's minor axis SB profile (inferred from resolved star counts) out to 65\,kpc, reaching $\mu_{V}\,{\simeq}\,33$\,\magsqarc\ --- among the deepest SB profiles ever measured. We measure a density slope of $-$2.59, consistent with the profile measured by the GHOSTS survey with \textit{HST} \citep{harmsen2017}. We also convert our star count profile to near-infrared SB and compare to \textit{WISE W1} measurements of the inner 10\,kpc of M81, finding good agreement. Using this calibrated SB profile, we estimate a total past accreted stellar mass for M81 of $1.16{\times}10^9\,M_{\odot}$\ --- indicating a largest past accretion of \textit{at most} the mass of the SMC. 
    \item M81's average $g{-}i$\ color profile out to 65\,kpc. We measure a flat color profile ($g{-}i\,{=}\,1.68$, [Fe/H] $\sim$ $-$1.2) from 20--60\,kpc, as seen by the GHOSTS survey \citep{monachesi2016a}. We also observe, for the first time, a steep negative color gradient ($\sim$0.05\,dex\,kpc$^{-1}$) at $R\,{=}$\,10--20\,kpc. Though we are unable to differentiate an accreted vs.~\textit{in situ} origin for the inner color gradient, M81's halo metallicity of [Fe/H]\,$\sim$\,$-$1.2 at 30\,kpc is in line with its past accreted mass of ${\sim}10^9\,M_{\odot}$, relative to the stellar halo mass--metallicity relation (see Figure \ref{fig:evolution}).
    \item Globally resolved stellar halo populations. Our metallicity-coded map of RGB stars reveals the triple interaction between M81, M82, and NGC 3077, highlighting the stark contrast between properties of M81's halo at large radii and the metal-rich debris around the interacting satellites.
    \item Stellar mass surface density on $\sim$1\,kpc scales, down to $\Sigma_{\star}\,{<}$\,10$^4\,M_{\odot}$\,kpc$^{-2}$. Using this sensitive stellar mass surface density map, we estimate the amount of tidal debris which is currently unbound from M82 and NGC 3077 --- ${\sim}5.4{\times}10^8\,M_{\odot}$, with an average metallicity of [Fe/H]\,$\sim$\,$-$0.9. This unbound debris represents a significant infusion of metal-rich material to the `current' stellar halo of M81. 
\end{enumerate}

Together, these measurements allow us to piece together `the saga of M81'. This MW-analog experienced a quiet history, accreting at most an SMC-mass satellite, likely sometime early in its life. Its current mergers with M82 and NGC 3077, however, have already altered M81's stellar halo properties on a short ($<$\,1\,Gyr) timescale, providing a substantial infusion of unbound metal-rich material. In the next several Gyrs, its merger with M82 will transform M81's halo from one of the least massive and metal-poorest, into one of the most massive and metal-rich halos known, rivaling (perhaps even exceeding) prototypical examples of massive halos such as that of M31. 

Furthermore, M81's stochastic stellar halo transition, from a low-mass and metal-poor halo to high-mass and metal-rich, is direct evidence that the \textit{diversity in stellar halo properties} at the MW-mass scale is primarily driven by \textit{diversity in the largest mergers} these galaxies have experienced. \\

We thank the anonymous referee for a careful and thorough review, which significantly improved this paper.

We thank Tom Jarrett and the WISE extragalactic working group for access to the \textit{WISE} surface brightness data for M81, as part of the \textit{WISE} Extended Source Catalog (WXSC). A.S. acknowledges support for this work by the National Science Foundation Graduate Research Fellowship Program under grant No.~DGE 1256260. Any opinions, findings, and conclusions or recommendations expressed in this material are those of the author(s) and do not necessarily reflect the views of the National Science Foundation. E.F.B. is partly supported by NASA grant NNG16PJ28C through subcontract from the University of Washington as part of the \textit{WFIRST} Infrared Nearby Galaxies Survey. AM acknowledges financial support from FONDECYT Regular 1181797 and funding from the Max Planck Society through a Partner Group grant.

Based on observations utilizing Pan-STARRS1 Survey. The Pan-STARRS1 Surveys (PS1) and the PS1 public science archive have been made possible through contributions by the Institute for Astronomy, the University of Hawaii, the Pan-STARRS Project Office, the Max-Planck Society and its participating institutes, the Max Planck Institute for Astronomy, Heidelberg and the Max Planck Institute for Extraterrestrial Physics, Garching, The Johns Hopkins University, Durham University, the University of Edinburgh, the Queen's University Belfast, the Harvard-Smithsonian Center for Astrophysics, the Las Cumbres Observatory Global Telescope Network Incorporated, the National Central University of Taiwan, the Space Telescope Science Institute, the National Aeronautics and Space Administration under Grant No. NNX08AR22G issued through the Planetary Science Division of the NASA Science Mission Directorate, the National Science Foundation Grant No. AST-1238877, the University of Maryland, Eotvos Lorand University (ELTE), the Los Alamos National Laboratory, and the Gordon and Betty Moore Foundation.

Based on observations obtained at the Subaru Observatory, which is operated by the National Astronomical Observatory of Japan, via the Gemini/Subaru Time Exchange Program. We thank the Subaru support staff --- particularly Akito Tajitsu, Tsuyoshi Terai, Dan Birchall, and Fumiaki Nakata --- for invaluable help preparing and carrying out the observing run. 

The authors wish to recognize and acknowledge the very significant cultural role and reverence that the summit of Maunakea has always had within the indigenous Hawaiian community. We are most fortunate to have the opportunity to conduct observations from this mountain.

\software{\texttt{HSC Pipeline} \citep{bosch2018}, \texttt{Matplotlib} \citep{matplotlib}, \texttt{NumPy} \citep{numpy-guide,numpy}, \texttt{Astropy} \citep{astropy}, \texttt{SciPy} \citep{scipy}, \texttt{SAOImage DS9} \citep{ds9}}

\bibliographystyle{aasjournal}

\begin{thebibliography}{}
\expandafter\ifx\csname natexlab\endcsname\relax\def\natexlab#1{#1}\fi
\providecommand{\url}[1]{\href{#1}{#1}}
\providecommand{\dodoi}[1]{doi:~\href{http://doi.org/#1}{\nolinkurl{#1}}}
\providecommand{\doeprint}[1]{\href{http://ascl.net/#1}{\nolinkurl{http://ascl.net/#1}}}
\providecommand{\doarXiv}[1]{\href{https://arxiv.org/abs/#1}{\nolinkurl{https://arxiv.org/abs/#1}}}

\bibitem[{{Astropy Collaboration} {et~al.}(2018){Astropy Collaboration},
  {Price-Whelan}, {Sip{\H{o}}cz}, {G{\"u}nther}, {Lim}, {Crawford}, {Conseil},
  {Shupe}, {Craig}, {Dencheva}, {Ginsburg}, {Vand erPlas}, {Bradley},
  {P{\'e}rez-Su{\'a}rez}, {de Val-Borro}, {Aldcroft}, {Cruz}, {Robitaille},
  {Tollerud}, {Ardelean}, {Babej}, {Bach}, {Bachetti}, {Bakanov}, {Bamford},
  {Barentsen}, {Barmby}, {Baumbach}, {Berry}, {Biscani}, {Boquien}, {Bostroem},
  {Bouma}, {Brammer}, {Bray}, {Breytenbach}, {Buddelmeijer}, {Burke},
  {Calderone}, {Cano Rodr{\'\i}guez}, {Cara}, {Cardoso}, {Cheedella}, {Copin},
  {Corrales}, {Crichton}, {D'Avella}, {Deil}, {Depagne}, {Dietrich}, {Donath},
  {Droettboom}, {Earl}, {Erben}, {Fabbro}, {Ferreira}, {Finethy}, {Fox},
  {Garrison}, {Gibbons}, {Goldstein}, {Gommers}, {Greco}, {Greenfield},
  {Groener}, {Grollier}, {Hagen}, {Hirst}, {Homeier}, {Horton}, {Hosseinzadeh},
  {Hu}, {Hunkeler}, {Ivezi{\'c}}, {Jain}, {Jenness}, {Kanarek}, {Kendrew},
  {Kern}, {Kerzendorf}, {Khvalko}, {King}, {Kirkby}, {Kulkarni}, {Kumar},
  {Lee}, {Lenz}, {Littlefair}, {Ma}, {Macleod}, {Mastropietro}, {McCully},
  {Montagnac}, {Morris}, {Mueller}, {Mumford}, {Muna}, {Murphy}, {Nelson},
  {Nguyen}, {Ninan}, {N{\"o}the}, {Ogaz}, {Oh}, {Parejko}, {Parley}, {Pascual},
  {Patil}, {Patil}, {Plunkett}, {Prochaska}, {Rastogi}, {Reddy Janga},
  {Sabater}, {Sakurikar}, {Seifert}, {Sherbert}, {Sherwood-Taylor}, {Shih},
  {Sick}, {Silbiger}, {Singanamalla}, {Singer}, {Sladen}, {Sooley},
  {Sornarajah}, {Streicher}, {Teuben}, {Thomas}, {Tremblay}, {Turner},
  {Terr{\'o}n}, {van Kerkwijk}, {de la Vega}, {Watkins}, {Weaver}, {Whitmore},
  {Woillez}, {Zabalza}, \& {Astropy Contributors}}]{astropy}
{Astropy Collaboration}, {Price-Whelan}, A.~M., {Sip{\H{o}}cz}, B.~M., {et~al.}
  2018, \aj, 156, 123, \dodoi{10.3847/1538-3881/aabc4f}

\bibitem[{{Bailin} {et~al.}(2011){Bailin}, {Bell}, {Chappell}, {Radburn-Smith},
  \& {de Jong}}]{bailin2011}
{Bailin}, J., {Bell}, E.~F., {Chappell}, S.~N., {Radburn-Smith}, D.~J., \& {de
  Jong}, R.~S. 2011, \apj, 736, 24, \dodoi{10.1088/0004-637X/736/1/24}

\bibitem[{{Barnes} \& {Hernquist}(1991)}]{barnes&hernquist1991}
{Barnes}, J.~E., \& {Hernquist}, L.~E. 1991, \apj, 370, L65,
  \dodoi{10.1086/185978}

\bibitem[{{Bell} {et~al.}(2017){Bell}, {Monachesi}, {Harmsen}, {de Jong},
  {Bailin}, {Radburn-Smith}, {D'Souza}, \& {Holwerda}}]{bell2017}
{Bell}, E.~F., {Monachesi}, A., {Harmsen}, B., {et~al.} 2017, \apj, 837, L8,
  \dodoi{10.3847/2041-8213/aa6158}

\bibitem[{{Bell} {et~al.}(2010){Bell}, {Xue}, {Rix}, {Ruhland}, \&
  {Hogg}}]{bell2010}
{Bell}, E.~F., {Xue}, X.~X., {Rix}, H.-W., {Ruhland}, C., \& {Hogg}, D.~W.
  2010, \aj, 140, 1850, \dodoi{10.1088/0004-6256/140/6/1850}

\bibitem[{{Bell} {et~al.}(2008){Bell}, {Zucker}, {Belokurov}, {Sharma},
  {Johnston}, {Bullock}, {Hogg}, {Jahnke}, {de Jong}, {Beers}, {Evans},
  {Grebel}, {Ivezi{\'c}}, {Koposov}, {Rix}, {Schneider}, {Steinmetz}, \&
  {Zolotov}}]{bell2008}
{Bell}, E.~F., {Zucker}, D.~B., {Belokurov}, V., {et~al.} 2008, \apj, 680, 295,
  \dodoi{10.1086/588032}

\bibitem[{{Bosch} {et~al.}(2018){Bosch}, {Armstrong}, {Bickerton}, {Furusawa},
  {Ikeda}, {Koike}, {Lupton}, {Mineo}, {Price}, {Takata}, {Tanaka}, {Yasuda},
  {AlSayyad}, {Becker}, {Coulton}, {Coupon}, {Garmilla}, {Huang}, {Krughoff},
  {Lang}, {Leauthaud}, {Lim}, {Lust}, {MacArthur}, {Mandelbaum}, {Miyatake},
  {Miyazaki}, {Murata}, {More}, {Okura}, {Owen}, {Swinbank}, {Strauss},
  {Yamada}, \& {Yamanoi}}]{bosch2018}
{Bosch}, J., {Armstrong}, R., {Bickerton}, S., {et~al.} 2018, Publications of
  the Astronomical Society of Japan, 70, S5, \dodoi{10.1093/pasj/psx080}

\bibitem[{{Bressan} {et~al.}(2012){Bressan}, {Marigo}, {Girardi}, {Salasnich},
  {Dal Cero}, {Rubele}, \& {Nanni}}]{bressan2012}
{Bressan}, A., {Marigo}, P., {Girardi}, L., {et~al.} 2012, \mnras, 427, 127,
  \dodoi{10.1111/j.1365-2966.2012.21948.x}

\bibitem[{{Bullock} \& {Johnston}(2005)}]{bullock&johnston2005}
{Bullock}, J.~S., \& {Johnston}, K.~V. 2005, \apj, 635, 931,
  \dodoi{10.1086/497422}

\bibitem[{{Bullock} {et~al.}(2001){Bullock}, {Kravtsov}, \&
  {Weinberg}}]{bullock2001}
{Bullock}, J.~S., {Kravtsov}, A.~V., \& {Weinberg}, D.~H. 2001, \apj, 548, 33,
  \dodoi{10.1086/318681}

\bibitem[{{Caldwell} {et~al.}(1998){Caldwell}, {Armandroff}, {Da Costa}, \&
  {Seitzer}}]{caldwell1998}
{Caldwell}, N., {Armandroff}, T.~E., {Da Costa}, G.~S., \& {Seitzer}, P. 1998,
  \aj, 115, 535, \dodoi{10.1086/300233}

\bibitem[{{Carollo} {et~al.}(2010){Carollo}, {Beers}, {Chiba}, {Norris},
  {Freeman}, {Lee}, {Ivezi{\'c}}, {Rockosi}, \& {Yanny}}]{carollo2010}
{Carollo}, D., {Beers}, T.~C., {Chiba}, M., {et~al.} 2010, \apj, 712, 692,
  \dodoi{10.1088/0004-637X/712/1/692}

\bibitem[{{Cautun} {et~al.}(2019){Cautun}, {Deason}, {Frenk}, \&
  {McAlpine}}]{cautun2019}
{Cautun}, M., {Deason}, A.~J., {Frenk}, C.~S., \& {McAlpine}, S. 2019, \mnras,
  483, 2185, \dodoi{10.1093/mnras/sty3084}

\bibitem[{{Chiboucas} {et~al.}(2013){Chiboucas}, {Jacobs}, {Tully}, \&
  {Karachentsev}}]{chiboucas2013}
{Chiboucas}, K., {Jacobs}, B.~A., {Tully}, R.~B., \& {Karachentsev}, I.~D.
  2013, \aj, 146, 126, \dodoi{10.1088/0004-6256/146/5/126}

\bibitem[{{Chiboucas} {et~al.}(2009){Chiboucas}, {Karachentsev}, \&
  {Tully}}]{chiboucas2009}
{Chiboucas}, K., {Karachentsev}, I.~D., \& {Tully}, R.~B. 2009, \aj, 137, 3009,
  \dodoi{10.1088/0004-6256/137/2/3009}

\bibitem[{{Conroy} {et~al.}(2019){Conroy}, {Naidu}, {Zaritsky}, {Bonaca},
  {Cargile}, {Johnson}, \& {Caldwell}}]{conroy2019}
{Conroy}, C., {Naidu}, R.~P., {Zaritsky}, D., {et~al.} 2019, arXiv e-prints,
  arXiv:1909.02007.
\newblock \doarXiv{1909.02007}

\bibitem[{{Covey} {et~al.}(2007){Covey}, {Ivezi{\'c}}, {Schlegel},
  {Finkbeiner}, {Padmanabhan}, {Lupton}, {Ag{\"u}eros}, {Bochanski}, {Hawley},
  {West}, {Seth}, {Kimball}, {Gogarten}, {Claire}, {Haggard}, {Kaib},
  {Schneider}, \& {Sesar}}]{covey2007}
{Covey}, K.~R., {Ivezi{\'c}}, {\v{Z}}., {Schlegel}, D., {et~al.} 2007, \aj,
  134, 2398, \dodoi{10.1086/522052}

\bibitem[{{Crnojevi{\'c}} {et~al.}(2016){Crnojevi{\'c}}, {Sand}, {Spekkens},
  {Caldwell}, {Guhathakurta}, {McLeod}, {Seth}, {Simon}, {Strader}, \&
  {Toloba}}]{crnojevic2016}
{Crnojevi{\'c}}, D., {Sand}, D.~J., {Spekkens}, K., {et~al.} 2016, \apj, 823,
  19, \dodoi{10.3847/0004-637X/823/1/19}

\bibitem[{{Davenport} {et~al.}(2014){Davenport}, {Ivezi{\'c}}, {Becker},
  {Ruan}, {Hunt-Walker}, {Covey}, {Lewis}, {AlSayyad}, \&
  {Anderson}}]{davenport2014}
{Davenport}, J. R.~A., {Ivezi{\'c}}, {\v{Z}}., {Becker}, A.~C., {et~al.} 2014,
  \mnras, 440, 3430, \dodoi{10.1093/mnras/stu466}

\bibitem[{{de Blok} {et~al.}(2008){de Blok}, {Walter}, {Brinks},
  {Trachternach}, {Oh}, \& {Kennicutt}}]{deblok2008}
{de Blok}, W.~J.~G., {Walter}, F., {Brinks}, E., {et~al.} 2008, \aj, 136, 2648,
  \dodoi{10.1088/0004-6256/136/6/2648}

\bibitem[{{de Blok} {et~al.}(2018){de Blok}, {Walter}, {Ferguson}, {Bernard},
  {van der Hulst}, {Neeleman}, {Leroy}, {Ott}, {Zschaechner}, {Zwaan}, {Yun},
  {Langston}, \& {Keating}}]{deblok2018}
{de Blok}, W.~J.~G., {Walter}, F., {Ferguson}, A. M.~N., {et~al.} 2018, \apj,
  865, 26, \dodoi{10.3847/1538-4357/aad557}

\bibitem[{{de Mello} {et~al.}(2008){de Mello}, {Smith}, {Sabbi}, {Gallagher},
  {Mountain}, \& {Harbeck}}]{deMello2008}
{de Mello}, D.~F., {Smith}, L.~J., {Sabbi}, E., {et~al.} 2008, \aj, 135, 548,
  \dodoi{10.1088/0004-6256/135/2/548}

\bibitem[{{Deason} {et~al.}(2011){Deason}, {Belokurov}, \&
  {Evans}}]{deason2011}
{Deason}, A.~J., {Belokurov}, V., \& {Evans}, N.~W. 2011, \mnras, 416, 2903,
  \dodoi{10.1111/j.1365-2966.2011.19237.x}

\bibitem[{{Deason} {et~al.}(2019){Deason}, {Belokurov}, \&
  {Sanders}}]{deason2019}
{Deason}, A.~J., {Belokurov}, V., \& {Sanders}, J.~L. 2019, \mnras, 2394,
  \dodoi{10.1093/mnras/stz2793}

\bibitem[{{D'Souza} \& {Bell}(2018{\natexlab{a}})}]{dsouza&bell2018a}
{D'Souza}, R., \& {Bell}, E.~F. 2018{\natexlab{a}}, \mnras, 474, 5300,
  \dodoi{10.1093/mnras/stx3081}

\bibitem[{{D'Souza} \& {Bell}(2018{\natexlab{b}})}]{dsouza&bell2018b}
---. 2018{\natexlab{b}}, Nature Astronomy, 2, 737,
  \dodoi{10.1038/s41550-018-0533-x}

\bibitem[{{Durrell} {et~al.}(2010){Durrell}, {Sarajedini}, \&
  {Chandar}}]{durrell2010}
{Durrell}, P.~R., {Sarajedini}, A., \& {Chandar}, R. 2010, \apj, 718, 1118,
  \dodoi{10.1088/0004-637X/718/2/1118}

\bibitem[{{Fattahi} {et~al.}(2019){Fattahi}, {Belokurov}, {Deason}, {Frenk},
  {G{\'o}mez}, {Grand }, {Marinacci}, {Pakmor}, \& {Springel}}]{fattahi2019}
{Fattahi}, A., {Belokurov}, V., {Deason}, A.~J., {et~al.} 2019, \mnras, 484,
  4471, \dodoi{10.1093/mnras/stz159}

\bibitem[{{Font} {et~al.}(2011){Font}, {McCarthy}, {Crain}, {Theuns}, {Schaye},
  {Wiersma}, \& {Dalla Vecchia}}]{font2011}
{Font}, A.~S., {McCarthy}, I.~G., {Crain}, R.~A., {et~al.} 2011, \mnras, 416,
  2802, \dodoi{10.1111/j.1365-2966.2011.19227.x}

\bibitem[{{Gallazzi} {et~al.}(2005){Gallazzi}, {Charlot}, {Brinchmann},
  {White}, \& {Tremonti}}]{gallazzi2005}
{Gallazzi}, A., {Charlot}, S., {Brinchmann}, J., {White}, S. D.~M., \&
  {Tremonti}, C.~A. 2005, \mnras, 362, 41,
  \dodoi{10.1111/j.1365-2966.2005.09321.x}

\bibitem[{{Gilbert} {et~al.}(2014){Gilbert}, {Kalirai}, {Guhathakurta},
  {Beaton}, {Geha}, {Kirby}, {Majewski}, {Patterson}, {Tollerud}, {Bullock},
  {Tanaka}, \& {Chiba}}]{gilbert2014}
{Gilbert}, K.~M., {Kalirai}, J.~S., {Guhathakurta}, P., {et~al.} 2014, \apj,
  796, 76, \dodoi{10.1088/0004-637X/796/2/76}

\bibitem[{{Gilbert} {et~al.}(2018){Gilbert}, {Tollerud}, {Beaton},
  {Guhathakurta}, {Bullock}, {Chiba}, {Kalirai}, {Kirby}, {Majewski}, \&
  {Tanaka}}]{gilbert2018}
{Gilbert}, K.~M., {Tollerud}, E., {Beaton}, R.~L., {et~al.} 2018, \apj, 852,
  128, \dodoi{10.3847/1538-4357/aa9f26}

\bibitem[{{Harmsen} {et~al.}(2017){Harmsen}, {Monachesi}, {Bell}, {de Jong},
  {Bailin}, {Radburn-Smith}, \& {Holwerda}}]{harmsen2017}
{Harmsen}, B., {Monachesi}, A., {Bell}, E.~F., {et~al.} 2017, \mnras, 466,
  1491, \dodoi{10.1093/mnras/stw2992}

\bibitem[{{High} {et~al.}(2009){High}, {Stubbs}, {Rest}, {Stalder}, \&
  {Challis}}]{high2009}
{High}, F.~W., {Stubbs}, C.~W., {Rest}, A., {Stalder}, B., \& {Challis}, P.
  2009, \aj, 138, 110, \dodoi{10.1088/0004-6256/138/1/110}

\bibitem[{{Hunter}(2007)}]{matplotlib}
{Hunter}, J.~D. 2007, Computing in Science and Engineering, 9, 90,
  \dodoi{10.1109/MCSE.2007.55}

\bibitem[{{Ibata} {et~al.}(2001){Ibata}, {Irwin}, {Lewis}, {Ferguson}, \&
  {Tanvir}}]{ibata2001}
{Ibata}, R., {Irwin}, M., {Lewis}, G., {Ferguson}, A. M.~N., \& {Tanvir}, N.
  2001, \nat, 412, 49.
\newblock \doarXiv{astro-ph/0107090}

\bibitem[{{Ibata} {et~al.}(2014){Ibata}, {Lewis}, {McConnachie}, {Martin},
  {Irwin}, {Ferguson}, {Babul}, {Bernard}, {Chapman}, {Collins}, {Fardal},
  {Mackey}, {Navarro}, {Pe{\~n}arrubia}, {Rich}, {Tanvir}, \&
  {Widrow}}]{ibata2014}
{Ibata}, R.~A., {Lewis}, G.~F., {McConnachie}, A.~W., {et~al.} 2014, \apj, 780,
  128, \dodoi{10.1088/0004-637X/780/2/128}

\bibitem[{{Ivezi{\'c}} {et~al.}(2007){Ivezi{\'c}}, {Smith}, {Miknaitis}, {Lin},
  {Tucker}, {Lupton}, {Gunn}, {Knapp}, {Strauss}, {Sesar}, {Doi}, {Tanaka},
  {Fukugita}, {Holtzman}, {Kent}, {Yanny}, {Schlegel}, {Finkbeiner},
  {Padmanabhan}, {Rockosi}, {Juri{\'c}}, {Bond}, {Lee}, {Stoughton}, {Jester},
  {Harris}, {Harding}, {Morrison}, {Brinkmann}, {Schneider}, \&
  {York}}]{ivezic2007}
{Ivezi{\'c}}, {\v{Z}}., {Smith}, J.~A., {Miknaitis}, G., {et~al.} 2007, \aj,
  134, 973, \dodoi{10.1086/519976}

\bibitem[{{Jang} {et~al.}(2020){Jang}, {de Jong}, {Holwerda}, {Monachesi},
  {Bell}, \& {Bailin}}]{jang2020}
{Jang}, I.~S., {de Jong}, R.~S., {Holwerda}, B.~W., {et~al.} 2020, \aap, 637,
  A8, \dodoi{10.1051/0004-6361/201936994}

\bibitem[{{Jarrett} {et~al.}(2003){Jarrett}, {Chester}, {Cutri}, {Schneider},
  \& {Huchra}}]{jarrett2003}
{Jarrett}, T.~H., {Chester}, T., {Cutri}, R., {Schneider}, S.~E., \& {Huchra},
  J.~P. 2003, \aj, 125, 525, \dodoi{10.1086/345794}

\bibitem[{{Jarrett} {et~al.}(2019){Jarrett}, {Cluver}, {Brown}, {Dale}, {Tsai},
  \& {Masci}}]{jarrett2019}
{Jarrett}, T.~H., {Cluver}, M.~E., {Brown}, M.~J.~I., {et~al.} 2019, arXiv
  e-prints, arXiv:1910.11793.
\newblock \doarXiv{1910.11793}

\bibitem[{{Jarrett} {et~al.}(2012){Jarrett}, {Masci}, {Tsai}, {Petty},
  {Cluver}, {Assef}, {Benford}, {Blain}, {Bridge}, \& {Donoso}}]{jarrett2012}
{Jarrett}, T.~H., {Masci}, F., {Tsai}, C.~W., {et~al.} 2012, \aj, 144, 68,
  \dodoi{10.1088/0004-6256/144/2/68}

\bibitem[{{Jarrett} {et~al.}(2013){Jarrett}, {Masci}, {Tsai}, {Petty},
  {Cluver}, {Assef}, {Benford}, {Blain}, {Bridge}, {Donoso}, {Eisenhardt},
  {Koribalski}, {Lake}, {Neill}, {Seibert}, {Sheth}, {Stanford}, \&
  {Wright}}]{jarrett2013}
---. 2013, \aj, 145, 6, \dodoi{10.1088/0004-6256/145/1/6}

\bibitem[{{Johnston} {et~al.}(2008){Johnston}, {Bullock}, {Sharma}, {Font},
  {Robertson}, \& {Leitner}}]{johnston2008}
{Johnston}, K.~V., {Bullock}, J.~S., {Sharma}, S., {et~al.} 2008, \apj, 689,
  936, \dodoi{10.1086/592228}

\bibitem[{{Kafle} {et~al.}(2012){Kafle}, {Sharma}, {Lewis}, \&
  {Bland-Hawthorn}}]{kafle2012}
{Kafle}, P.~R., {Sharma}, S., {Lewis}, G.~F., \& {Bland-Hawthorn}, J. 2012,
  \apj, 761, 98, \dodoi{10.1088/0004-637X/761/2/98}

\bibitem[{{Karachentsev} {et~al.}(2013){Karachentsev}, {Makarov}, \&
  {Kaisina}}]{karachentsev2013}
{Karachentsev}, I.~D., {Makarov}, D.~I., \& {Kaisina}, E.~I. 2013, \aj, 145,
  101, \dodoi{10.1088/0004-6256/145/4/101}

\bibitem[{{Karachentsev} {et~al.}(2006){Karachentsev}, {Dolphin}, {Tully},
  {Sharina}, {Makarova}, {Makarov}, {Karachentseva}, {Sakai}, \&
  {Shaya}}]{karachentsev2006}
{Karachentsev}, I.~D., {Dolphin}, A., {Tully}, R.~B., {et~al.} 2006, \aj, 131,
  1361, \dodoi{10.1086/500013}

\bibitem[{{Karachentseva} \&
  {Karachentsev}(1998)}]{karachentseva&karachentsev1998}
{Karachentseva}, V.~E., \& {Karachentsev}, I.~D. 1998, \aaps, 127, 409,
  \dodoi{10.1051/aas:1998109}

\bibitem[{{King}(1962)}]{king1962}
{King}, I. 1962, \aj, 67, 471, \dodoi{10.1086/108756}

\bibitem[{{Lancaster} {et~al.}(2019){Lancaster}, {Koposov}, {Belokurov},
  {Evans}, \& {Deason}}]{lancaster2019}
{Lancaster}, L., {Koposov}, S.~E., {Belokurov}, V., {Evans}, N.~W., \&
  {Deason}, A.~J. 2019, \mnras, 486, 378, \dodoi{10.1093/mnras/stz853}

\bibitem[{{Lim} {et~al.}(2013){Lim}, {Hwang}, \& {Lee}}]{lim2013}
{Lim}, S., {Hwang}, N., \& {Lee}, M.~G. 2013, \apj, 766, 20,
  \dodoi{10.1088/0004-637X/766/1/20}

\bibitem[{{Mackereth} \& {Bovy}(2019)}]{mackereth2019}
{Mackereth}, J.~T., \& {Bovy}, J. 2019, arXiv e-prints, arXiv:1910.03590.
\newblock \doarXiv{1910.03590}

\bibitem[{{Mackey} {et~al.}(2019){Mackey}, {Lewis}, {Brewer}, {Ferguson},
  {Veljanoski}, {Huxor}, {Collins}, {C{\^o}t{\'e}}, {Ibata}, {Irwin}, {Martin},
  {McConnachie}, {Pe{\~n}arrubia}, {Tanvir}, \& {Wan}}]{mackey2019}
{Mackey}, D., {Lewis}, G.~F., {Brewer}, B.~J., {et~al.} 2019, \nat, 574, 69,
  \dodoi{10.1038/s41586-019-1597-1}

\bibitem[{{Magnier} {et~al.}(2013){Magnier}, {Schlafly}, {Finkbeiner}, {Juric},
  {Tonry}, {Burgett}, {Chambers}, {Flewelling}, {Kaiser}, {Kudritzki},
  {Morgan}, {Price}, {Sweeney}, \& {Stubbs}}]{magnier2013}
{Magnier}, E.~A., {Schlafly}, E., {Finkbeiner}, D., {et~al.} 2013, The
  Astrophysical Journal Supplement Series, 205, 20,
  \dodoi{10.1088/0067-0049/205/2/20}

\bibitem[{{Martin} {et~al.}(2013){Martin}, {Ibata}, {McConnachie}, {Mackey},
  {Ferguson}, {Irwin}, {Lewis}, \& {Fardal}}]{martin2013}
{Martin}, N.~F., {Ibata}, R.~A., {McConnachie}, A.~W., {et~al.} 2013, \apj,
  776, 80, \dodoi{10.1088/0004-637X/776/2/80}

\bibitem[{{McConnachie}(2012)}]{mcconnachie2012}
{McConnachie}, A.~W. 2012, \aj, 144, 4, \dodoi{10.1088/0004-6256/144/1/4}

\bibitem[{{McConnachie} {et~al.}(2018){McConnachie}, {Ibata}, {Martin},
  {Ferguson}, {Collins}, {Gwyn}, {Irwin}, {Lewis}, {Mackey}, {Davidge},
  {Arias}, {Conn}, {C{\^o}t{\'e}}, {Crnojevic}, {Huxor}, {Penarrubia},
  {Spengler}, {Tanvir}, {Valls-Gabaud}, {Babul}, {Barmby}, {Bate}, {Bernard},
  {Chapman}, {Dotter}, {Harris}, {McMonigal}, {Navarro}, {Puzia}, {Rich},
  {Thomas}, \& {Widrow}}]{mcconnachie2018}
{McConnachie}, A.~W., {Ibata}, R., {Martin}, N., {et~al.} 2018, \apj, 868, 55,
  \dodoi{10.3847/1538-4357/aae8e7}

\bibitem[{{Merritt} {et~al.}(2016){Merritt}, {van Dokkum}, {Abraham}, \&
  {Zhang}}]{merritt2016}
{Merritt}, A., {van Dokkum}, P., {Abraham}, R., \& {Zhang}, J. 2016, \apj, 830,
  62, \dodoi{10.3847/0004-637X/830/2/62}

\bibitem[{{Monachesi} {et~al.}(2016{\natexlab{a}}){Monachesi}, {Bell},
  {Radburn-Smith}, {Bailin}, {de Jong}, {Holwerda}, {Streich}, \&
  {Silverstein}}]{monachesi2016a}
{Monachesi}, A., {Bell}, E.~F., {Radburn-Smith}, D.~J., {et~al.}
  2016{\natexlab{a}}, \mnras, 457, 1419, \dodoi{10.1093/mnras/stv2987}

\bibitem[{{Monachesi} {et~al.}(2016{\natexlab{b}}){Monachesi}, {G{\'o}mez},
  {Grand}, {Kauffmann}, {Marinacci}, {Pakmor}, {Springel}, \&
  {Frenk}}]{monachesi2016b}
{Monachesi}, A., {G{\'o}mez}, F.~A., {Grand}, R. J.~J., {et~al.}
  2016{\natexlab{b}}, \mnras, 459, L46, \dodoi{10.1093/mnrasl/slw052}

\bibitem[{{Monachesi} {et~al.}(2013){Monachesi}, {Bell}, {Radburn-Smith},
  {Vlaji{\'c}}, {de Jong}, {Bailin}, {Dalcanton}, {Holwerda}, \&
  {Streich}}]{monachesi2013}
{Monachesi}, A., {Bell}, E.~F., {Radburn-Smith}, D.~J., {et~al.} 2013, \apj,
  766, 106, \dodoi{10.1088/0004-637X/766/2/106}

\bibitem[{{Monachesi} {et~al.}(2019){Monachesi}, {G{\'o}mez}, {Grand },
  {Simpson}, {Kauffmann}, {Bustamante}, {Marinacci}, {Pakmor}, {Springel}, \&
  {Frenk}}]{monachesi2019}
{Monachesi}, A., {G{\'o}mez}, F.~A., {Grand }, R. J.~J., {et~al.} 2019, \mnras,
  485, 2589, \dodoi{10.1093/mnras/stz538}

\bibitem[{{Notni} {et~al.}(2004){Notni}, {Karachentsev}, \&
  {Makarova}}]{notni2004}
{Notni}, P., {Karachentsev}, I.~D., \& {Makarova}, L.~N. 2004, Astronomische
  Nachrichten, 325, 307, \dodoi{10.1002/asna.200310175}

\bibitem[{{Okamoto} {et~al.}(2015){Okamoto}, {Arimoto}, {Ferguson}, {Bernard},
  {Irwin}, {Yamada}, \& {Utsumi}}]{okamoto2015}
{Okamoto}, S., {Arimoto}, N., {Ferguson}, A. M.~N., {et~al.} 2015, \apj, 809,
  L1, \dodoi{10.1088/2041-8205/809/1/L1}

\bibitem[{Oliphant(2006)}]{numpy-guide}
Oliphant, T.~E. 2006, A guide to NumPy, Vol.~1 (Trelgol Publishing USA)

\bibitem[{{Pillepich} {et~al.}(2015){Pillepich}, {Madau}, \&
  {Mayer}}]{pillepich2015}
{Pillepich}, A., {Madau}, P., \& {Mayer}, L. 2015, \apj, 799, 184,
  \dodoi{10.1088/0004-637X/799/2/184}

\bibitem[{{Querejeta} {et~al.}(2015){Querejeta}, {Meidt}, {Schinnerer},
  {Cisternas}, {Mu{\~n}oz-Mateos}, {Sheth}, {Knapen}, {van de Ven}, {Norris},
  {Peletier}, {Laurikainen}, {Salo}, {Holwerda}, {Athanassoula}, {Bosma},
  {Groves}, {Ho}, {Gadotti}, {Zaritsky}, {Regan}, {Hinz}, {Gil de Paz},
  {Menendez-Delmestre}, {Seibert}, {Mizusawa}, {Kim}, {Erroz-Ferrer}, {Laine},
  \& {Comer{\'o}n}}]{querejeta2015}
{Querejeta}, M., {Meidt}, S.~E., {Schinnerer}, E., {et~al.} 2015, \apjs, 219,
  5, \dodoi{10.1088/0067-0049/219/1/5}

\bibitem[{{Radburn-Smith} {et~al.}(2011){Radburn-Smith}, {de Jong}, {Seth},
  {Bailin}, {Bell}, {Brown}, {Bullock}, {Courteau}, {Dalcanton}, {Ferguson},
  {Goudfrooij}, {Holfeltz}, {Holwerda}, {Purcell}, {Sick}, {Streich}, {Vlajic},
  \& {Zucker}}]{radburn-smith2011}
{Radburn-Smith}, D.~J., {de Jong}, R.~S., {Seth}, A.~C., {et~al.} 2011, The
  Astrophysical Journal Supplement Series, 195, 18,
  \dodoi{10.1088/0067-0049/195/2/18}

\bibitem[{{Rodr{\'\i}guez-Merino} {et~al.}(2011){Rodr{\'\i}guez-Merino},
  {Rosa-Gonz{\'a}lez}, \& {Mayya}}]{rodriguez-merino2011}
{Rodr{\'\i}guez-Merino}, L.~H., {Rosa-Gonz{\'a}lez}, D., \& {Mayya}, Y.~D.
  2011, \apj, 726, 51, \dodoi{10.1088/0004-637X/726/1/51}

\bibitem[{{Sabbi} {et~al.}(2008){Sabbi}, {Gallagher}, {Smith}, {de Mello}, \&
  {Mountain}}]{sabbi2008}
{Sabbi}, E., {Gallagher}, J.~S., {Smith}, L.~J., {de Mello}, D.~F., \&
  {Mountain}, M. 2008, \apjl, 676, L113, \dodoi{10.1086/587548}

\bibitem[{{Schlafly} \& {Finkbeiner}(2011)}]{schlafly&finkbeiner2011}
{Schlafly}, E.~F., \& {Finkbeiner}, D.~P. 2011, \apj, 737, 103,
  \dodoi{10.1088/0004-637X/737/2/103}

\bibitem[{{Sheth} {et~al.}(2010){Sheth}, {Regan}, {Hinz}, {Gil de Paz},
  {Men{\'e}ndez-Delmestre}, {Mu{\~n}oz-Mateos}, {Seibert}, {Kim},
  {Laurikainen}, {Salo}, {Gadotti}, {Laine}, {Mizusawa}, {Armus},
  {Athanassoula}, {Bosma}, {Buta}, {Capak}, {Jarrett}, {Elmegreen},
  {Elmegreen}, {Knapen}, {Koda}, {Helou}, {Ho}, {Madore}, {Masters},
  {Mobasher}, {Ogle}, {Peng}, {Schinnerer}, {Surace}, {Zaritsky},
  {Comer{\'o}n}, {de Swardt}, {Meidt}, {Kasliwal}, \& {Aravena}}]{sheth2010}
{Sheth}, K., {Regan}, M., {Hinz}, J.~L., {et~al.} 2010, \pasp, 122, 1397,
  \dodoi{10.1086/657638}

\bibitem[{{Smercina} {et~al.}(2017){Smercina}, {Bell}, {Slater}, {Price},
  {Bailin}, \& {Monachesi}}]{smercina2017}
{Smercina}, A., {Bell}, E.~F., {Slater}, C.~T., {et~al.} 2017, \apj, 843, L6,
  \dodoi{10.3847/2041-8213/aa78fa}

\bibitem[{{Smithsonian Astrophysical Observatory}(2000)}]{ds9}
{Smithsonian Astrophysical Observatory}. 2000, {SAOImage DS9: A utility for
  displaying astronomical images in the X11 window environment}.
\newblock \doeprint{0003.002}

\bibitem[{{Spitzer} \& {Shapiro}(1972)}]{spitzer&shapiro1972}
{Spitzer}, Lyman, J., \& {Shapiro}, S.~L. 1972, \apj, 173, 529,
  \dodoi{10.1086/151442}

\bibitem[{{Streich} {et~al.}(2014){Streich}, {de Jong}, {Bailin}, {Goudfrooij},
  {Radburn-Smith}, \& {Vlajic}}]{streich2014}
{Streich}, D., {de Jong}, R.~S., {Bailin}, J., {et~al.} 2014, \aap, 563, A5,
  \dodoi{10.1051/0004-6361/201220956}

\bibitem[{{Toomre} \& {Toomre}(1972)}]{toomre&toomre1972}
{Toomre}, A., \& {Toomre}, J. 1972, \apj, 178, 623, \dodoi{10.1086/151823}

\bibitem[{Van Der~Walt {et~al.}(2011)Van Der~Walt, Colbert, \&
  Varoquaux}]{numpy}
Van Der~Walt, S., Colbert, S.~C., \& Varoquaux, G. 2011, Computing in Science
  \& Engineering, 13, 22

\bibitem[{{Virtanen} {et~al.}(2020){Virtanen}, {Gommers}, {Oliphant},
  {Haberland}, {Reddy}, {Cournapeau}, {Burovski}, {Peterson}, {Weckesser},
  {Bright}, {van der Walt}, {Brett}, {Wilson}, {Millman}, {Mayorov}, {Nelson},
  {Jones}, {Kern}, {Larson}, {Carey}, {Polat}, {Feng}, {Moore}, {Vand erPlas},
  {Laxalde}, {Perktold}, {Cimrman}, {Henriksen}, {Quintero}, {Harris},
  {Archibald}, {Ribeiro}, {Pedregosa}, {van Mulbregt}, \& {SciPy 1. 0
  Contributors}}]{scipy}
{Virtanen}, P., {Gommers}, R., {Oliphant}, T.~E., {et~al.} 2020, Nature
  Methods, 17, 261, \dodoi{10.1038/s41592-019-0686-2}

\bibitem[{{von Hoerner}(1957)}]{vonHorner1957}
{von Hoerner}, S. 1957, \apj, 125, 451, \dodoi{10.1086/146321}

\bibitem[{{Watkins} {et~al.}(2016){Watkins}, {Mihos}, \&
  {Harding}}]{watkins2016}
{Watkins}, A.~E., {Mihos}, J.~C., \& {Harding}, P. 2016, \apj, 826, 59,
  \dodoi{10.3847/0004-637X/826/1/59}

\bibitem[{{White} \& {Rees}(1978)}]{white&rees1978}
{White}, S.~D.~M., \& {Rees}, M.~J. 1978, \mnras, 183, 341,
  \dodoi{10.1093/mnras/183.3.341}

\bibitem[{{Williams} {et~al.}(2015){Williams}, {Dalcanton}, {Bell}, {Gilbert},
  {Guhathakurta}, {Dorman}, {Lauer}, {Seth}, {Kalirai}, {Rosenfield}, \&
  {Girardi}}]{williams2015}
{Williams}, B.~F., {Dalcanton}, J.~J., {Bell}, E.~F., {et~al.} 2015, \apj, 802,
  49, \dodoi{10.1088/0004-637X/802/1/49}

\bibitem[{{Yun} {et~al.}(1994){Yun}, {Ho}, \& {Lo}}]{yun1994}
{Yun}, M.~S., {Ho}, P.~T.~P., \& {Lo}, K.~Y. 1994, \nat, 372, 530,
  \dodoi{10.1038/372530a0}

\bibitem[{{Zolotov} {et~al.}(2009){Zolotov}, {Willman}, {Brooks}, {Governato},
  {Brook}, {Hogg}, {Quinn}, \& {Stinson}}]{zolotov2009}
{Zolotov}, A., {Willman}, B., {Brooks}, A.~M., {et~al.} 2009, \apj, 702, 1058,
  \dodoi{10.1088/0004-637X/702/2/1058}

\end{thebibliography}

\appendix 
\restartappendixnumbering

\section{Minor Axis Profile Table}
In Table \ref{tab:minax} we provide the radial profiles along M81's minor axis for $\mu_{V}$\ ($V$-band SB) and average $g{-}i$\ color, respectively. See Figure \ref{fig:minax-sb} \& \ref{fig:minax-color} for plots of each profile.

\newcolumntype{p}{!{\extracolsep{50pt}}c!{\extracolsep{0pt}}}

\begin{table}[!h]
\centering
\hspace{-1.5cm}\begin{tabular}{ p p p }
\multicolumn{3}{ p }{\textbf{Table A1.} Minor Axis SB \& Color Profiles} \\
\hline
$R$ & $\mu_{V}$ & $g{-}i$ \\
(kpc) & (\magsqarc) & (mag) \vspace{3pt} \\
\hline
10 & $<$\,28.74 & 1.89$^{{-}0.04}_{{+}0.04}$ \\ 
12 & $<$\,28.74 & 1.88$^{{-}0.03}_{{+}0.03}$ \\ 
14 & 28.70\,$\pm$\,0.28 & 1.81$^{{-}0.04}_{{+}0.04}$ \\ 
16 & 29.28\,$\pm$\,0.19 & 1.78$^{{-}0.04}_{{+}0.04}$ \\ 
18 & 29.47\,$\pm$\,0.17 & 1.70$^{{-}0.05}_{{+}0.05}$ \\ 
20 & 29.94\,$\pm$\,0.13 & 1.74$^{{-}0.06}_{{+}0.06}$ \\ 
22 & 30.40\,$\pm$\,0.11 & 1.67$^{{-}0.07}_{{+}0.07}$ \\ 
24 & 30.77\,$\pm$\,0.10 & 1.68$^{{-}0.07}_{{+}0.07}$ \\ 
26 & 30.95\,$\pm$\,0.10 & 1.69$^{{-}0.08}_{{+}0.08}$ \\ 
28 & 31.27\,$\pm$\,0.10 & 1.66$^{{-}0.08}_{{+}0.08}$ \\ 
30 & 31.36\,$\pm$\,0.09 & 1.65$^{{-}0.09}_{{+}0.09}$ \\ 
32 & 31.48\,$\pm$\,0.09 & 1.69$^{{-}0.09}_{{+}0.09}$ \\ 
34 & 31.62\,$\pm$\,0.09 & 1.65$^{{-}0.09}_{{+}0.09}$ \\ 
36 & 31.57\,$\pm$\,0.09 & 1.66$^{{-}0.09}_{{+}0.09}$ \\ 
38 & 31.87\,$\pm$\,0.10 & 1.71$^{{-}0.10}_{{+}0.10}$ \\ 
40 & 31.79\,$\pm$\,0.07 & 1.69$^{{-}0.06}_{{+}0.06}$ \\ 
45 & 31.91\,$\pm$\,0.07 & 1.68$^{{-}0.06}_{{+}0.06}$ \\ 
50 & 32.32\,$\pm$\,0.07 & 1.70$^{{-}0.06}_{{+}0.06}$ \\ 
55 & 32.76\,$\pm$\,0.08 & 1.69$^{{-}0.07}_{{+}0.07}$ \\ 
60 & 32.76\,$\pm$\,0.08 & 1.67$^{{-}0.06}_{{+}0.06}$ \\ 
65 & 32.99\,$\pm$\,0.08 & 1.67$^{{-}0.07}_{{+}0.07}$ 
\vspace{3pt} \\ 
\hline
\end{tabular}
\begin{minipage}{0.6\textwidth}
\caption{The radial minor axis average surface brightness and average $g{-}i$\ color profiles as shown in Figure \ref{fig:minax-sb} \& \ref{fig:minax-color}. See \S\,\ref{sec:sb-prof} and \S\,\ref{sec:color-prof} for discussion of how the measurements and uncertainties are computed.}\label{tab:minax}
\end{minipage}
\end{table}

\end{document}